\documentclass[aps,prb,twocolumn,superscriptaddress,floatfix]{revtex4}
\usepackage{amsmath,amssymb}
\usepackage{bm}
\usepackage{graphicx}

\newcommand{\beq}{\begin{equation}}
\newcommand{\eeq}{\end{equation}}

\newcommand{\fc}{\mbox{\scriptsize FC}}
\newcommand{\tf}{\mbox{\scriptsize TF}}
\newcommand{\fl}{\mbox{\scriptsize FL}}
\newcommand{\qcp}{\mbox{\scriptsize QCP}}
\newcommand{\crit}{\mbox{\scriptsize cr}}

\begin{document}
\title{Adaptation of the Landau-Migdal Quasiparticle Pattern to
Strongly Correlated Fermi Systems}

\author{V.~A.~Khodel}
\affiliation{Russian Research Centre Kurchatov
Institute, Moscow, 123182, Russia}
\affiliation{McDonnell Center for the Space Sciences \&
Department of Physics, Washington University,
St.~Louis, MO 63130, USA}
\author{J.~W.~Clark}
\affiliation{McDonnell Center for the Space Sciences \&
Department of Physics, Washington University,
St.~Louis, MO 63130, USA}
\author{M.~V.~Zverev}
\affiliation{Russian Research Centre Kurchatov
Institute, Moscow, 123182, Russia}
\affiliation{Moscow Institute of Physics and Technology, Dolgoprudnyi, Moscow region, 141700, Russia}

\date{\today}

\begin{abstract}

A quasiparticle pattern advanced in Landau's first article on
Fermi liquid theory is adapted to elucidate the properties of
a class of strongly correlated Fermi systems characterized by a
Lifshitz phase diagram featuring a quantum critical point (QCP)
where the density of states diverges. The necessary condition
for stability of the Landau Fermi Liquid state is shown to break
down in such systems, triggering a cascade of topological phase
transitions that lead, without symmetry violation, to states
with multi-connected Fermi surfaces. The end point of this
evolution is found to be an exceptional state whose spectrum of
single-particle excitations exhibits a completely flat portion
at zero temperature.  Analysis of the evolution of the temperature
dependence of the single-particle spectrum yields results that
provide a natural explanation of classical behavior of this class
of Fermi systems in the QCP region.

\end{abstract}


\maketitle

\section{Preamble}

Once upon a time at a traditional meeting of nestlings of the
Migdal school, A.\ B.\ began his speech with words that stunned
the audience.

{\it Until recent times I was proud that I have never published
an incorrect article. But is it actually a true cause for pride?
Can an experienced mountaineer take pride in not having broken
any ribs, or a professional motorcyclist, that his legs are still
intact?  The correct answer is no!  It just means that these people may
have run to the best of their abilities, but certainly not more.}

This memorial article, dedicated to a great physicist and a great man,
is devoted to a problem first discussed around 20 years ago.\cite{ks,vol,noz}
 The cited works considered the possibility
of a breakdown of the conventional Landau-Migdal quasiparticle
pattern \cite{lan1,lan2,migt,migqp} of phenomena observed in Fermi
liquids (FL), associated specifically with rearrangement of the $T=0$
Landau quasiparticle momentum distribution   $n_F(p)=\theta(p_F-p)$.

Over the last decade, experimental studies of non-Fermi liquid (NFL)
behavior of strongly correlated systems have extended the frontiers
of low-temperature condensed matter physics.\cite{godfrin1995,
godfrin1996,godfrin1998,saunders1,saunders2,coleman,lohr,steglich}
During the same period, a number of theorists have engaged in
efforts to extend the frontiers of FL theory with the aim of
explaining this anomalous behavior.\cite{physrep,zb,shagp,zkb,
jpcm2004,shaghf,prb2005,yak,shagrev, khodel2007,prb2008,jetplett2009,
shagrep2010}

There is a famous Migdal Correspondence Criterion for judging
new theories, which boils down to this:

{\it First an foremost, the proposed theory must be able to match
results of previous, well-tested theoretical descriptions. It
is only of secondary importance that it matches the relevant
experimental data. Why so? Because as a rule, the experiments
of burning interest were performed yesterday, and therefore the
results obtained may be flawed, whereas theoretical physics
stands as body of knowledge created and honed by a multitude
over three hundred years.}

The extended quasiparticle picture to be reviewed and discussed here
meets the Migdal Criterion. It reduces naturally to the the standard FL
picture when dealing with conventional Fermi liquids. At the same time,
convincing experimental evidence supporting this extension of the
Landau-Migdal vision, while present, remains scarce---possibly because
of the short time the relevant experimental programs have been in
operation.\cite{godfrin1996,loh,aronson,lohr}  Even so, one cause
for the lack of unambiguous correlations between available experimental data and corresponding theoretical results may be conceptual (or technical) errors made by the theorists in developing the new theory, errors that are yet to be exposed. In this case, we should not be too upset, since according to Migdal's provocative challenge to his assembled group, such errors tell us that the struggling theorists, being professionals, did, after all, run beyond the breaking point of their abilities.

\section{Routes to breakdown of standard Fermi-Liquid theory}

Any theory has own limits of applicability, and FL theory is no exception.
Conventionally these limits are imputed to violation of Pomeranchuk
stability conditions.\cite{pom} Such violation is associated with
second-order phase transitions that are accompanied with jumps of
the specific heat $C(T)$ and cusps of the spin susceptibility $\chi(T)$.
 Upon approach to the transition point, spontaneous creation and
enhancement of fluctuations suppresses the value of the $z$ factor
that determines the quasiparticle weight in the single-particle
state at the Fermi surface.  At the transition point, $z$ vanishes,
signaling a breakdown of the FL quasiparticle picture.\cite{doniach,dyugaev}

A different domain where predictions of FL theory prove to be fallacious
has been discovered and explored during the last decade. This is the
regime of the so-called quantum critical point (QCP) revealed in
experimental studies of 2D liquid $^3$He and heavy-fermion metals.
In this domain, the specific heat $C(T)$ and spin susceptibility
$\chi(T)$, both proportional to the density of states $N(T)$ (and
hence to the effective mass $M^*$), are found to diverge as the
temperature $T$ goes to zero.\cite{godfrin1995,godfrin1996,
godfrin1998,saunders1,saunders2,coleman,lohr,steglich}

The presence of a QCP is the hallmark that distinguishes strongly
correlated Fermi systems from those with weak or moderate correlations.
In this respect, three-dimensional (3D) liquid $^3$He, for which
$M^*$ remains finite at any density, belongs to the class of systems
with moderate correlations.  Its 2D counterpart, on the other hand,
is assigned to the class of strongly correlated Fermi systems based
on evidence for the existence of a QCP provided by experimental
studies of dense $^3$He films.\cite{godfrin1995,godfrin1996,godfrin1998,
saunders1,saunders2}

It is well to emphasize that the divergence of the specific heat
$C(T)$ that is observed in the QCP region has little in common with
the jumps in $C(T)$ inherent in second-order phase transitions.
Notwithstanding this fact, theoretical explanations of the failure
of FL theory in the QCP region have commonly linked its apparent
breakdown with attendant second-order phase transitions.\cite{doniach,dyugaev,hertz,millis,coleman} However, as they occur in the QCP region, such transitions are found to be quite atypical. In most cases, even the structure of the order parameters remains unknown, and the properties of states beyond the QCP defy explanation within the standard scaling theory of second-order phase transitions.
One is led to conclude that the viability of such transitions as triggers of the observed NFL behavior is problematic.

In this article, we pursue another explanation of NFL behavior
in the QCP region, attributing the breakdown of standard FL theory
to violation of the {\it necessary stability condition} for the
Landau state.\cite{physrep}  The salient feature of the proposed
scenario is rearrangement of single-particle degrees of freedom
driven by {\it topological} phase transitions (TPT's). This
mechanism stands in direct contrast to the violation of
{\it sufficient} Pomeranchuk stability conditions,\cite{pom}
which entail rearrangement of {\it collective degrees of freedom}
in second-order phase transitions.

Since the topological scenario does not implicate the violation
of any Pomeranchuk stability condition, the catastrophic suppression
of the quasiparticle weight posited in the collective scenario
for the QCP does not take place, and failure of standard FL theory
in the QCP region must have another explanation. Indeed, the
fundamental FL formula
\beq
\epsilon(p)=v_F(p-p_F)\equiv p_F(p-p_F)/M^*
\label{flspec}
\eeq
for the single-particle spectrum $\epsilon(p)$ measured from the
Fermi surface, becomes powerless when the effective mass $M^*$
diverges. If such a divergence is present, additional terms of the
Taylor expansion of $\epsilon(p)$ must be added to the right side
of Eq.~(\ref{flspec}) (see Ref.~\onlinecite{prb2005}), giving rise
to a change of sign of the Fermi velocity. This in its turn
signals a violation of the necessary stability condition for
the conventional Landau state (as seen in Sec.~V.B),
resulting in a {\it topological} rearrangement of the Fermi
surface that necessitates modification of the standard FL
formalism.

\section{Reprise of the standard FL quasiparticle picture}

To set the stage for discussion of the required modifications,
we recall that the heart of the Landau-Migdal quasiparticle
picture is the postulate that there exists a one-to-one
correspondence between the totality of real, decaying
single-particle excitations of the actual Fermi liquid and a
system of immortal interacting quasiparticles. There are two
facets of this correspondence.  First, the number of quasiparticles
is equal to the given number $N$ of particles, a condition expressed
as
\beq
{\rm Tr}\int n({\bf p})\,d\upsilon={N\over V}\equiv \rho,
\label{part}
\eeq
where $n({\bf p})$ is the quasiparticle momentum distribution,
$\rho$ is the density, ${\rm Tr}$ implies summation over spin and
isospin variables, and $d\upsilon=d{\bf p}/(2\pi)^D$ is a volume
element in a momentum space of dimension $D$.  Second, all
thermodynamic quantities, notably the ground state energy $E$ and
entropy $S$, are treated as functionals of the quasiparticle
momentum distribution $n({\bf p})$. In particular, the entropy
$S$ of the real system is given by the ideal-Fermi-gas combinatorial expression
\beq
{S\over V}=-{\rm Tr}\int [n({\bf p})\ln n({\bf p})
+(1-n({\bf p}))\ln (1-n({\bf p}))]d\upsilon.
\label{entr}
\eeq
In homogeneous matter, a standard variational procedure based on
Eq.~(\ref{entr}) and involving restrictions that impose conservation
of the particle number and energy, leads to the result \cite{lan1}
\beq
n(p)=\left[ 1+e^{\epsilon(p)/T}\right]^{-1}.
\label{dist}
\eeq
While this relation resembles the corresponding Fermi-Dirac formula
for the ideal Fermi gas, the quasiparticle energy $\epsilon(p)$
measured from the chemical potential does not coincide with the
bare single-particle energy $\epsilon^0_p=p^2/(2M)-\mu$.
Instead, it is given by the variational derivative
\beq
\epsilon(p) =\delta \Omega/\delta n(p)\
\label{spec}
\eeq
of the thermodynamic potential $\Omega=E-\mu N$.

It is significant that the most seminal article cited by Landau
in his pioneering works \cite{lan1,lan2} was authored by
A.\ B.\ Migdal,\cite{migj} who had advanced an idea that allows
unambiguous determination of the Fermi momentum $p_F$ of the
quasiparticle system.  In his numerous papers and books, A.\ B.\
also adapted the Landau quasiparticle concept to finite systems
where the phenomenological spectrum (\ref{flspec}) becomes
more complicated due to its dependence on the shape of the
self-consistent field acting on a quasiparticle in an inhomogeneous
medium. Indeed, his famous book \cite{migt} completed the
creation of the FL formulation through the exposition of a
quantitative theory of the behavior of Fermi liquids in external
fields. This book has become a touchstone for modern theoretical
description of the properties of atomic nuclei.  Moreover, together
with A.\ I.\ Larkin he developed a quantitative theory of superfluid
Fermi liquids.\cite{migl} It is in recognition of these splendid
achievements that the FL quasiparticle theory is often referred to
as the Landau-Migdal quasiparticle pattern.

In principle, the quasiparticle spectrum $\epsilon(p)$ can be
evaluated by means of the relation \cite{lan1,lan2,lanl,trio}
\beq
v({\bf p})={\partial\epsilon(p)\over\partial {\bf p}} =
  {{\bf p}\over M} + \int\! f({\bf p},{\bf p_1})\,
  {\partial n(p_1)\over\partial {\bf p_1}}\, d\upsilon_1,
\label{lansp}
\eeq
stemming from the assertion that the single-particle energy
$\epsilon(p)$ is a functional of the quasiparticle momentum
distribution $n(p)$.  Alternatively, this equation may be derived
from the Galilean invariance of the system Hamiltonian and gauge
invariance.\cite{pit,yaf2001} From this coincidence we infer that
the spectrum $\epsilon(p)$ is, indeed, a functional of $n$, thereby
reinforcing the validity of the basic Landau postulate that
the ground state energy $E$ and other thermodynamic quantities
can be treated as functionals of $n$.

Eq.(\ref{lansp}) provides a nonlinear integral equation for
self-consistent determination of the quasiparticle spectrum $\epsilon(p)$
and the quasiparticle momentum distribution (\ref{dist}), treating
the Landau interaction function $f({\bf p},{\bf p}_1)$ as
phenomenological input.  It is worth noting that the correct relation
between the function $f({\bf p},{\bf p}_1)$ and the scattering
amplitude $\Gamma$ was first established by Migdal.\cite{comment}
This function turns out to be the so-called $\omega$-limit of
$\Gamma({\bf p},{\bf p}_1,{\bf k},\omega)$, where the energy
transfer $\omega$ and momentum transfer $k$ both tend to zero, but
in such a way that $k/\omega\to 0$.  In this limit, only regular
Feynman diagrams contribute to $f$ in full force; consequently
the interaction function $f$ cannot be evaluated within the FL
formalism itself.  Accordingly, $f$ is commonly specified by
a set of phenomenological parameters, namely the harmonics of
its Legendre polynomial expansion.

Following Landau (cf.\ Eq.~(4) in Ref.~\onlinecite{lan1}), one
postulates that at $T=0$, solutions of Eq.~(\ref{lansp}) arrange
themselves in such a way that the Fermi velocity $v_F$ {\it always}
keeps a positive value. Necessarily, then, the quasiparticle
momentum distribution $n(p,T=0)$ coincides with the Fermi step
 $n_F(p)=\theta(p_F-p)$
 appropriate to a noninteracting Fermi gas.
In turn, the relation (\ref{part}) reduces to the famous
Landau-Luttinger theorem, expressed simply as
\beq
\rho={p^3_F\over 3\pi^2}
\label{ll}
\eeq
for the case of 3D homogeneous matter. The resulting quasiparticle
picture, in which a system of fermions is treated as a ``gas of
interacting quasiparticles'' \cite{migt} works flawlessly for
conventional Fermi liquids, including 3D liquid $^3$He, the
matter inside atomic nuclei, and the electron liquid in
alkali metals, where the correlations are moderate or weak.
However, as already indicated, this picture begins to fail
upon entry into the QCP region, where the effective mass
$M^*$ diverges.

\section{Theoretical alternatives for divergence of the effective mass}

Discrepancies between FL predictions and experimental data were first
observed in films of liquid $^3$He at relatively high areal density.
In stark contrast to the ordinary FL behavior exhibited by the bulk
liquid, the 2D data for the spin susceptibility $\chi(T)$ and
Sommerfeld ratio $\gamma=C(T)/T$ of specific heat $C(T)$ to
temperature $T$ soar upward as $T\to 0$,\cite{godfrin1995,godfrin1996,
godfrin1998,saunders1,saunders2} whereas both quantities must
become constant in FL theory. Analogous behavior has been discovered
and documented in strongly correlated electron systems of
solids \cite{coleman,lohr,steglich} and in the 2D electron gas as
realized in MOSFETs.\cite{pudalov,kravchenko,shashkin1,shashkin2,
shashrev}

Such a behavior is ascribed to divergence of the effective mass $M^*$,
first revealed theoretically in microscopic calculations of the
single-particle spectrum $\epsilon(p)$ of the dilute 3D homogeneous
electron gas \cite{ks1,zks} (for the 2D case, see Ref.~\onlinecite{bz}).
Two different sources for divergence of $M^*$ in nonsuperfluid Fermi
systems can be identified in terms of the textbook formula
\beq
{M\over M^*}=z\left[1+\left({\partial\Sigma(p,\varepsilon)
\over\partial\epsilon^0_p}\right)_0\right],
\label{mefft}
\eeq
where the quasiparticle weight $z$ in the single-particle state is
given by $z=[1-\left(\partial\Sigma(p,\varepsilon)/
\partial\varepsilon\right)_0]^{-1}$, with $\Sigma$ representing
the mass operator. Here and henceforth, the subscript $0$ indicates
that a function of $p$ and $\varepsilon$ is evaluated at the Fermi
surface.

As seen from this formula, one way in which a divergence of
$M^*$ could arise (first discussed by Doniach and Engelsberg
\cite{doniach}) is through divergence of the derivative
$\left(\partial\Sigma(p,\varepsilon)/\partial\varepsilon\right)_0$
at a critical density $\rho_c$ where some second-order phase
transition occurs.  Another option invokes {\it a change of sign}
of the sum $1+\left(\partial\Sigma(p,\varepsilon)/
\partial\epsilon^0_p\right)_0$ at a critical density $\rho_{\infty}$.
Correspondingly, there are two different scenarios for the QCP.
In the first scenario---the most popular until recently---it is
the energy dependence of the self-energy $\Sigma(p,\varepsilon)$
that plays the decisive role. In this {\it collective scenario},
the posited divergence of
$\left(\partial\Sigma(p,\varepsilon)/\partial\varepsilon\right)_0$
at the points of second-order phase transitions results in
a vanishing renormalization factor $z$, and the QCP is identified
with the end point of the line $T_N(\rho)$ traced
 in the phase diagram by a second-order
phase transition. The essence of this scenario is captured in
the maxim \cite{coleman} that in the vicinity of any second-order
phase transition, ``Quasiparticles get heavy and die.''

In the alternative {\it topological} scenario, it is instead the momentum dependence of the mass operator $\Sigma(p,\varepsilon)$ that gives rise to the QCP at a critical density $\rho_{\infty}$ where the sum $1+\left(\partial\Sigma(p,\varepsilon)/\partial\epsilon^0_p\right)_0$ changes sign.
Let us recall that on the disordered side of the QCP, system properties {\it obey standard FL theory}. This fact, as we will see, is
incompatible with the collective scenario. Indeed, divergence of the effective mass imposes some restriction on the first harmonic $f_1$ entering Eq.~(\ref{lansp}). By way of illustration, we address a familiar 3D case in which simple manipulations performed on Eq.~(\ref{lansp})
at the critical QCP density $\rho_{\infty}=p^3_{\infty}/3\pi^2$ yield
\beq
{v_F(\rho_{\infty})\over v^0_F}\equiv {M\over M^*(\rho_{\infty})}
=1-{1\over 3}  F^0_1(\rho_{\infty})=0,
\label{rel01}
\eeq
where $v^0_F=p_F/M$.

In principle, given the bare interaction potential $V$, the harmonic $f_1$ can be evaluated within a microscopic approach to FL theory (for examples, see
Refs.~\onlinecite{physrep,feenberg}),
in which this quantity is a continuous functions of the density $\rho$.  Then the left-hand side of Eq.~(\ref{rel01})
must change sign beyond the point where the relation (\ref{rel01})
is met, implying that the Fermi velocity $v_F$, evaluated with the Landau quasiparticle momentum distribution $n_F(p)$, {\it necessarily} changes sign at the QCP. This behavior conflicts with the collective scenario for the QCP, in which such a sign change is impossible. We conclude that if FL theory is applicable on the disordered side of the QCP, the collective scenario is not relevant to the QCP.
The sign change results in the  violation of the necessary stability condition for the Landau state. Nevertheless,  as we will see in Sec.V.B, the original Landau quasiparticle  picture holds on both sides of the QCP, implying that in contrast to the collective scenario, the QCP itself and the line $T_N(\rho)$ of putatively associated second-order phase transitions are {\it separated} from each other.

In some influential theoretical articles devoted to the physics
of the QCP, it has been claimed that switching on the interactions
between particles fails to produce a significant momentum dependence
in the effective interaction function $f$, and hence that the
topological mechanism is irrelevant. This assertion cannot withstand
scrutiny. The natural measure of the strength of momentum-dependent
forces in the medium is provided by the dimensionless first harmonic
of the interaction function $f({\bf p},{\bf p}_1)$ of Landau theory.
To avoid any misunderstandings it is worth replacing the divergent effective mass $M^*$ appearing in the dimensionless Landau harmonic $F_1=f_1p_FM^*/\pi^2$ by the free-particle mass to obtain $F^0_1=f_1p_FM/\pi^2$ as a relevant characteristics of the strength of  the momentum dependence of  the effective interactions.
In a system such as 3D liquid $^3$He
where the correlations are of moderate strength, the result
$F^0_1\geq 2.0$ for this measure extracted from specific-heat
data is already rather large.  The data on 2D liquid $^3$He are
yet more damaging to the claim of minimal momentum dependence,
since the effective mass is found to {\it diverge} in dense films,\cite{godfrin1995,godfrin1996,godfrin1998,saunders1,saunders2}
implying that $F^0_1>3.0$. In considering the occurrence of QCP
phenomena in strongly correlated systems of {\it ionic} crystals,
it should be borne in mind that the electron effective mass
is greatly enhanced due to electron-phonon interactions that
subserve polaron effects.\cite{pekar,alex,alexk}

The irrelevance of the collective scenario may be drawn with respect to the notion that  a Pomeranchuk instability drives the QCP. To be definite, the argument may be framed in terms of a ferromagnetic phase transition, for which the associated Pomeranchuk stability condition reads
\beq
1+G_0(\rho_{\infty})\equiv
1+g_0(\rho_{\infty}){p_{\infty}M^*(\rho_{\infty})\over \pi^2}=0.
     \label{rel00}
\eeq
Here we set $\rho=\rho_{\infty}$ because in the collective scenario,
the point of divergence of the effective mass coincides with
the point of a second-order phase transition.

The crucial point is that at the QCP, both the relations (\ref{rel01}) and
(\ref{rel00}) must be met {\it simultaneously}.
Thus if at the QCP, the Pomeranchuk stability condition is violated then the right-hand sides of both Eq.~(\ref{rel00}) and Eq.~(\ref{rel01}) must vanish at the same density $\rho_{\infty}$. However, this vanishing
 can only happen {\it accidentally} or with the aid of an external magnetic
field. Otherwise, we encounter a dilemma: either $F^0_1(\rho_{\infty})=3$
and hence the effective mass diverges while the spin susceptibility
remains finite; or else $1+G_0(\rho_{\infty})=0$, and the spin
susceptibility diverges while the effective mass remains finite.
Anyway, in both the cases, the end point of the line $T_N(\rho)$ is separated from the QCP. Such a separation of phases was first
observed in 2D liquid  $^3$He \cite{saunders2} and was also recently
documented in precision experiments on heavy-fermion metals.\cite{bud'ko,stegcol}

In studies of the class of heavy-fermion metals, the information most
valuable for an understanding of their properties is drawn from studies
of those compounds in which localized magnetic moments
of ions forming the crystal lattice interact with narrow electron bands
located near the Fermi level.  The full theory of this so-called
Kondo lattice problem is far from complete (for example, see the
monograph of Hewson.\cite{hewson}) During the last decade
certain results from this theory were applied in the analysis of
the quantum critical phenomena in heavy-fermion metals.\cite{si,zhu,steglich,
pnac,pepin2010} In the corresponding scenario for the QCP, called
the Kondo breakdown model, the Kondo resonance is destroyed at
the QCP, causing the Kondo temperature $T_K$ to vanish and the
FL picture to be recovered on the disordered side of the QCP.
Quantitatively, this model, with the dynamical exponent ${\cal Z}= 3$,\cite{pepin2010} differs from the standard spin-density wave (SDW) model where ${\cal Z}= 2$, 
providing more realistic critical indexes. Notwithstanding this fact, the Kondo breakdown scenario for the QCP is burdened by the same shortcomings as the SDW scenario.
In all, the weight of the above considerations compels us to ``place the cart and the horse'' in developing a viable theory of the QCP, that is, to shift the focus from collective mechanisms to topological rearrangement of the Fermi surface
as the origin of NFL behavior and the key player in the QCP regime.

\section{Emergence of flat bands as a universal feature
of strongly correlated Fermi systems}

In standard FL theory, applicable to weakly and moderately correlated
Fermi systems, the $T=0$ ground-state momentum quasiparticle distribution
$n(p)$ is the Fermi step   $n_F(p)=\theta(p_F-p)$.
In this case, solution of the Landau relation (\ref{rel01}) for the quasiparticle spectrum $\epsilon(p)$ is obviated, being reduced to a simple momentum integration. The solutions for $\epsilon(p)$ prove to be continuous functions,
with all derivatives existing everywhere in momentum space. On the
other hand, in strongly correlated Fermi systems exhibiting a QCP,
the basic FL formula (\ref{flspec}) becomes fallacious, together with
the asserted property $n(p)=n_F(p)$. As we shall see, the
{\it necessary} condition for stability of the Landau state
is {\it inevitably} violated at some point in the QCP domain,
implying that the minimum of the functional $E[n]$ is displaced
from the Fermi step $n_F(p)$ to another point in the admissible
space of momentum distributions specified by the Pauli restriction
$0\leq n(p) \leq 1$.

\subsection{Case study: Rearrangement of the Landau state in a simple model}

In practice, numerical solution of the basic relation (\ref{lansp})
should demonstrate whether the true $T=0$ ground-state momentum
distribution $n_*(p)$ coincides with the Fermi step $n_F(p)$ or
with something more interesting.  However, we need not resort to
cumbersome numerical calculations to elucidate the rearrangement
of the Landau state in strongly correlated Fermi systems. To
illustrate this point, consider as an example
a simple model where the energy functional $E(n)$ has the form
\begin{eqnarray}
E(n)&=&{\rm Tr}
\int {p^2\over 2M}n(p) d\upsilon \nonumber \\
&+& {1\over 2}
{\rm Tr}_1 {\rm Tr}_2 \!\!
\int
n(p_1)n(p_2) f({\bf p}_1-{\bf p}_2) d\upsilon_1d\upsilon_2 \qquad
\label{lanfe}
\end{eqnarray}
that results in the following relation
\beq
\epsilon(p)={p^2\over 2M}-\mu+\int f({\bf p}-{\bf p}_1)n(p_1) d\upsilon_1
\label{rel1}
\eeq
between the quasiparticle momentum distribution $n(p)$ and the single-particle spectrum $\epsilon(p)$. In what follows we focus on key factors that promote a rearrangement of the Landau state as the correlations gain strength, assuming the interaction function  to be as follows: $f(q)=\lambda/q$, with the zero and first harmonics coinciding with each other: $f_0=f_1=\lambda/p_F$. It is this form of $f$ that was employed in the first article \cite{ks} where the phenomenon of flattening of single-particle spectra  was uncovered.
According to Ref.~\onlinecite{ks}, the necessary stability condition for the Landau state is violated at the critical value
\beq
F^0_1(\rho_{\infty})={f_1p_FM\over \pi^2}\equiv { M\lambda_{\qcp}\over \pi^2}=3  \  .
\label{rel102}
\eeq
Comparison of this result with  Eq.~(\ref{rel01}) shows that the effective mass $M^*$ does diverge at this point.

A straightforward estimate for the FL ground-state energy per particle
is obtained by inserting the Fermi step $n_F$ into Eq.~(\ref{lanfe}) to yield
\beq
{E(n_F)\over N}= {3p^2_F\over 10M}\left(1+{2F^0_1\over 3}\right),
\label{efl}
\eeq
If $F^0_1$ is large, the interaction term is dominant.  To diminish this term and thereby lower $E(n)$, the momentum distribution must be spread as much as possible,
until this is rendered counterproductive due to the consequent increase of the kinetic term.

To estimate the effect of a rearrangement from $n(p)$ to $n_*(p)$, let
us assume the new ground-state momentum distribution takes the form
$n_*(p)=\nu\, \theta(p_f-p)$, with $\nu<1$. Particle-number
conservation then implies $p_f = \nu^{-1/3} p_F$, and the kinetic
energy is increased by a factor $(p_f/p_F)^2$.  If the interaction
function $f$ were momentum-independent, no rearrangement would occur.
In the model posed, however, $f(q)$ falls off as $1/q$,
 and therefore the replacement $n_F(p)\to n_*(p)$ leads to the following result
\beq
{E(n_*)\over N}
= {3p^2_F\over 10M}\left[{p^2_f \over p^2_F}
+{2p_F\over 3 p_f}F^0_1 \right].
\label{moden}
\eeq
This energy estimate is minimized with respect to the parameter
$p_f$ by the ratio
\beq
{p_f\over p_F}=\left({F^0_1\over 3}\right)^{1/3},
\eeq
thus fixing $\nu$. This ratio must exceed unity, lest the Pauli restriction be violated; hence we surmise that the results obtained are meaningful provided $F^0_1>3$.  Referring to Eq.~(\ref{rel01}), we then infer that if this inequality is met, the Fermi velocity $v_F$ does change its sign, and the Landau state is indeed destabilized. With these results, the energy advantage of the rearranged state, $\Delta E_*=E(n_*)-E(n_F)$, is readily estimated:
\beq
\Delta E_*= -{3p^2_F\over 10M}\,
\left[\left({F^0_1\over 3}\right)^{1/3}-1\right]^2
\left[1+2\left({F^0_1\over 3}\right)^{1/3}\right] \ .
\eeq
Since the variational procedure is performed on a wider class of
solutions, the rearrangement ensures a {\it negative} supplement
$\Delta E_*$.

Naturally, one cannot clarify all details of the rearrangement of
the Landau state by means of simple variational estimates. It will
be seen below that in correlated Fermi systems where the interaction
function is strongly repulsive, the momentum distribution $n_*$
that minimizes the functional $E(n)$ does in fact dive inside
the definitional domain $D$ of $n(p)$, rather than switching
between one boundary point of this domain and the other. In this case,
the true ground state momentum distribution $n_*(p)$ is found as
a solution of the variational equation \cite{ks}
\beq
{\delta E(n(p))\over \delta n(p)}-\mu =0.
\label{varfl}
\eeq

\begin{figure}[t]
\includegraphics[width=0.78\linewidth,height=0.62\linewidth]{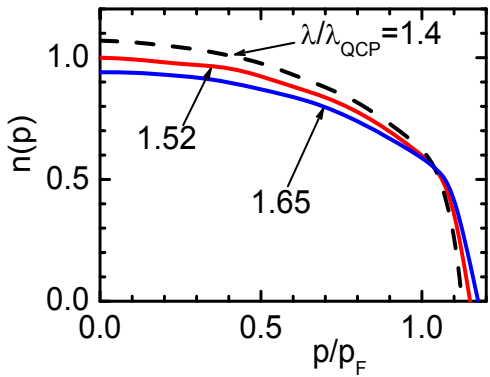}
\caption{Quasiparticle filling $n(p)$ calculated for $\kappa=0.07\,p_F$ and different values of coupling constant $\lambda$ in the model with interaction $f(q)=\lambda (q^2+\kappa^2)^{-1}$.}
\label{fig:FC_reg}
\end{figure}

As a concrete illustration, Fig.~\ref{fig:FC_reg} presents results
from numerical calculations performed in Ref.~\onlinecite{zkb} for
the quasiparticle filling determined from Eq.~(\ref{varfl}),
employing $f(q)=\lambda (q^2+\kappa^2)^{-1}$ with $\kappa=0.07\,p_F$.
The resulting quasiparticle momentum distribution $n_*(p)$, smooth
over the whole region $0<p<p_f$, is found by means of a regularization
procedure that is effective in treating ill-posed problems.  As
applied here, this procedure involves the introduction of a
very small additional local interaction between quasiparticles
of the form $U_0\delta({\bf p}-{\bf p}_1)$.  Eq.~(\ref{varfl})
then has the specific expression
\beq
{p^2\over 2M}+U_0n_*(p)+\lambda \int{n_*(p_1)\over({\bf p}-{\bf p}_1)^2+\kappa^2}\,d\upsilon_1-\mu=0,
\label{FC_regul}
\eeq
and was solved by a conventional grid method, taking
$U_0/\varepsilon_F^0=10^{-3}$. As seen in Fig.~\ref{fig:FC_reg},
a smooth solution of Eq.~(\ref{FC_regul}) exists for every
$\lambda$ value.  However, {\it this solution does not satisfy
the Pauli principle} until $\lambda$ attains a critical value
$\lambda_{\crit}=1.52\,\lambda_{\qcp}$, where $\lambda_{\qcp}$
corresponds to the first topological transition associated with
the QCP.  At $\lambda>\lambda_{\crit}$, the Pauli principle becomes
immaterial.  In the illustrative case considered, the Fermi volume
is found to expand or swell by as much as 30\%.

\subsection{Flattening of the spectrum of single-particle excitations
in strongly correlated Fermi systems}

One striking feature of the phenomenon of rearrangement of the
Landau state at $\lambda>\lambda_{\crit}$ is a profound flattening
of the spectrum of single-particle excitations $\epsilon(p)$.
The emergence of the flat band can be illuminated without
elaborate calculations. The flattening of $\epsilon(p)$ stems
from the necessary condition for stability of the ground
state,\cite{physrep} which requires that the change of the
ground-state energy $E_0$ produced by any admissible variation
of the momentum distribution $n(p)$ be nonnegative
\beq
\delta E_0=\int \epsilon(p;n_F)\delta n(p)d\upsilon \geq 0.
\label{necs0}
\eeq
Starting from the Landau state, the admissible variations of $n(p)$
from $n_F(p)$ must obey the Pauli principle, being negative at
$p<p_F$ and positive at $p>p_F$. In the standard FL theory, the signs
of $\epsilon(p)$ and $p-p_F$ always coincide and therefore the
Landau state is stable.  However, when we address the rearranged
filling of momentum states represented by $n_*(p)$, the quasiparticle
occupancies differ from the FL values 0 and 1 in an interval
$[p_i,p_f]$; accordingly the admissible variations of $n_*(p)$
in this range do not have a definite sign. The only way to avoid
violation of the inequality (\ref{necs0}) in this situation is
to make the spectrum $\epsilon(p)$ {\it completely flat} in
the domain $ p_i<p<p_f$.  This phenomenon, originally called
fermion condensation \cite{ks} and later swelling the Fermi
surface,\cite{noz} or emergence of flat bands, was discovered
20 years ago.  It was recently rediscovered by Lee \cite{lee}
in a different context, investigating the finite-charge-density
sector of conformal field theory (CFT) based on the AdS/CFT
gravity/gauge duality.

\subsection{Breakdown of the analyticity of solutions of the
Landau equation for the spectrum $\epsilon(p)$}

From the mathematical standpoint, the FL functional
$E_{\fl}(n)$
is equivalent to any functional of the Thomas-Fermi (TF) family
of functionals that depend on the density $\rho({\bf r})$,
even though $E_{\fl}(n)$ and $E_{\tf}(\rho)$ refer to different
definitional domains of the arguments $n$ and $\rho$. For the
ensuing development, it is convenient to write $E_{\tf}(\rho)$ in
the generalized form
\begin{eqnarray}
E_{\tf}(\rho)
&=&\tau(\rho(r)) +\int\rho(r)U(r)d{\bf r} \nonumber \\
   &+& {1\over 2}\int\!\!\int \rho({\bf r}_1)
\rho({\bf r}_2)V({\bf r}_1-{\bf r}_2)d{\bf r}_1 d{\bf r}_2,
\label{tf}
\end{eqnarray}
with the normalization condition
\beq
\int \rho(r)d{\bf r}=Z,
\label{normz}
\eeq
where the {\it finite} number $Z$ is the atomic charge.  In
Eq.~(\ref{tf}), we identify $\tau(\rho)$ as the electron kinetic
energy, $U(r)$ as an external field, and $V(r)$ as the two-particle
interaction, vanishing for $r \to \infty$.  The minimum of the
functional (\ref{tf}) is determined from the standard variational
condition
\beq
\epsilon(r)\equiv {\delta E_{\tf}(\rho)\over\delta\rho(r)}-\mu =0.
\label{varcl}
\eeq
The solution $\rho(r)$ of this equation must be everywhere nonnegative;
in addition, it must vanish at $r\to \infty$ to meet the normalization
condition (\ref{normz}).  Explicitly, one finds
\beq
{p^2_F(r)\over 2M} -Z{e^2\over r}+e^2\int {\rho(r_1)
\over |{\bf r}-{\bf r}_1|} d{\bf r}_1-\mu=0,
\label{varcle}
\eeq
where $p_F(r)=(3\pi^2\rho(r))^{1/3}$.

The density $\rho(r)$ is known to decay rapidly as $r$ increases
beyond the TF radius $\simeq 1/(Me^2)$, implying that the integral on the
left side of Eq.~(\ref{varcle}) and the external field cancel
each other at distances greatly exceeding this radius.  Since $\mu$
vanishes in a neutral atom, the asymptotic conditions are met,
and there exists a solution $\rho(r)$ that is everywhere
continuous, with all its derivatives defined.\cite{LL5}

The preceding argument breaks down for the case of an ionized atom.
This is evident already from the singly-ionized case: all terms on
the left in Eq.~(\ref{varcle}) vanish as $r \to \infty$ {\it but one,}
namely the chemical potential $\mu$, which has a finite value.\cite{LL5}
Thus, Eq.~(\ref{varcle}) cannot be satisfied at $r\to \infty$. We
conclude that this equation only applies within a {\it finite domain}
defined by $r<R$, with $R$ treated as the boundary of the ionized
atom beyond which $\rho(r)$ vanishes identically.\cite{LL5} The
equations determining the minimum of the functional (\ref{tf}) now
become
\begin{eqnarray}
{p^2_F(r)\over 2M} &-&Z{e^2\over r} +e^2\int\limits_0^R
{\rho(r_1)\over |{\bf r}-{\bf r}_1|}d{\bf r}_1-\mu=0, \quad  r\leq R,
\nonumber\\
\rho(r)&\equiv& 0, \quad r>R.
\label{varm}
\end{eqnarray}
Having obtained the corresponding density distribution, the
function $\epsilon(r)$ of Eq.~(\ref{varcl}) is evaluated
in closed form as
\begin{eqnarray}
\epsilon(r)&=&0 \  , \quad r\leq R, \nonumber\\
\epsilon(r)&=&-Z{e^2\over r}+e^2\int\limits_0^R {\rho(r_1)\over
|{\bf r}-{\bf r}_1|} d{\bf r}_1-\mu, \quad r>R.
\end{eqnarray}
The energy $\epsilon(r)$ vanishes identically only at $r\leq R$,
whereas at $r>R$ it changes somewhat with the observation point $r$.
Since any analytic function that vanishes identically in some
domain must be identically zero everywhere, we see that analytic
solutions of the TF variational problem {\it simply do not exist}
in the case of an ionized atom.

In the preceding development, we have demonstrated that
the Thomas-Fermi problem is ill posed, by witnessing the
breakdown of analyticity due to the impossibility of reconciling
(i) the variational equation that is to determine the
solution by functional minimization with (ii) the prescribed
asymptotic behavior of the desired solution.   Similarly,
the task of finding a satisfactory minimum of the Landau
functional $E_{\fl}(n)$ also belongs to the class of ill-posed
problems.  Since the first term in Eq.~(\ref{FC_regul}) diverges
at $p\to \infty$, there are no solutions of the Landau
variational problem that are continuous with derivatives
everywhere in momentum space.  Therefore the domain of solution
must be restricted, as in the TF problem (though in momentum
space rather than coordinate space) leading to the following
equation analogous to first of Eqs.~(\ref{varm})
\beq
{p^2\over  2M}+\int f({\bf p}-{\bf p}_1) n_*(p_1)
d\upsilon_1-\mu=0, \quad  p\in {\cal C}.
\eeq
A NFL solution $n_*(p)$ continuous in the momentum domain $\cal C$
shows {\it no Migdal jump} at the Fermi momentum $p=p_F$.
Correspondingly, in this domain
the basic property $n^2(p)=n(p)$ obeyed by the standard $T=0$
FL quasiparticle momentum distribution {\it no longer holds}.
Outside the domain ${\cal C}$, the spectrum $\epsilon(p)$ ceases
to be flat, and accordingly the distribution $n_*(p)$ coincides
with the Fermi step, being 1 for occupied states and 0, for
empty ones.  As stated above, the phenomenon of ``swelling''
of the Fermi surface implies that it expands from a line to
a surface in 2D, and from a surface to a volume in 3D.  In
other words, beyond the point of fermion condensation, the
single-particle spectrum $\epsilon(p)$ acquires a completely
flat portion $\epsilon(p) \equiv 0$ in the momentum interval
$p_i<p<p_f$ (the so-called {\it fermion condensate} (FC)).

With the aid of the variational condition (\ref{varfl}), one
can understand that the system with the FC {\it cannot} be
treated as a gas of interacting quasiparticles, because in
case a particle with momentum ${\bf p}\in {\cal C}$ is
added to the system, the addition results in a rearrangement
of the {\it whole} distribution function $n_*(p)$ in the
${\cal C}$ region.  This is an essential aspect of the
modification of the FL quasiparticle picture required to describe
strongly correlated Fermi systems.

\subsection{Systems hosting a FC as a special class of Fermi liquids}

The flattening of the spectrum $\epsilon(p)$ in systems with a FC
is reflected in the structure of the single-particle Green function
$G(p,\varepsilon)$ on the imaginary axes in the limit
$\varepsilon\to 0$, where \cite{vol}
\beq
G(p,\varepsilon)={1\over \varepsilon}, \quad p\in {\cal C}.
\label{volg}
\eeq
This makes the essential difference in evaluation of the
{\it topological charge}, given in the 2D case by the integral
\cite{vol,volrev}
\beq
{\cal N}=\int\limits_{\gamma} \left(G(p,\varepsilon)\right)^{-1}\, \partial_l
G(p,\varepsilon) {dl\over 2\pi i},  \quad p\in {\cal C},
\label{voln}
\eeq
where the Green function is considered on the imaginary energy axis
$\varepsilon=i\xi$ and the integration is performed over a
{\it small} contour $\gamma$ in the $({\bf p},\xi)$-space embracing
the Fermi line.  For conventional Fermi liquids and states with
a multi-connected Fermi surface, the topological charge ${\cal N}$
is an integer, whereas for a states having a completely flat portion
in the spectrum $\epsilon(p)$, integration over the contour embracing
the Fermi line yields a {\it half-odd-integral} value of
${\cal N}$.\cite{vol}  Accordingly, from the topological point
of view, systems hosting a FC form a {\it special class} of Fermi liquids.

In dealing with the whole Fermi-surface region, Eq.~(\ref{volg})
needs a slight generalization.  To explain this modification,
we introduce a minute temperature $T$.  It is then natural to
write the FC Green function in the form
\begin{equation}
G(p,\varepsilon,T)= {1-n_*( p,T)\over \varepsilon{+}i\gamma}+
{n_*( p,T)\over \varepsilon{-}i\gamma},  \quad p\in {\cal C},
\label{green}
\end{equation}
familiar in low-temperature physics, the occupation numbers
$n_*(p,T)$ being given by the Landau formula (\ref{dist}). In
writing this formula the ratio $\gamma/\varepsilon$ is assumed
to be being much less than unity, which applies at least for systems
having a small proportion of FC.

\subsection{Low-temperature expansion in systems with a FC}

It is a key feature of the phenomenon of fermion condensation
that a tiny elevation of temperature has practically no effect
on the $T=0$ momentum distribution, so that Eq.~(\ref{dist}) can
be inverted to obtain \cite{noz}
\begin{equation}
\epsilon(p,T\to 0)= T\ln {1-n_*(p)\over n_*(p)}, \qquad p\in {\cal C}.
\label{spte}
\end{equation}
This means that at extremely low temperatures, the dispersion of the
single-particle spectrum $\epsilon(p)$ in the FC domain is
proportional to $T$, thus allowing one to identify (or redefine)
the FC as the totality of single-particle states having a
linear-in-$T$ dispersion of their associated spectrum $\epsilon(p,T)$.
   As a matter of fact, the formula (\ref{spte}) is instrumental to
developing low-temperature expansions of the basic quantities
$\epsilon(p,T)$ and $n(p,T)$.
 To clarify how such expansions are generated, we
 first decompose the quasiparticle momentum distribution $n(p,T)$
as follows:
\beq
n(p,T)=n_*(p)+T\nu (p,T),
\label{dec}
\eeq
where $n_*(p)$ is the $T=0$ solution of Eq.~(\ref{rel1}). To facilitate the analysis we assume that the FC occupies that part of momentum space lying between $p=0$ to $p=p_f$, so that $n_*(p)$ vanishes identically at $p\geq p_f$.

Upon inserting the decomposition (\ref{dec}) into Eq.~(\ref{rel1})
and replacing there $\epsilon(p)$ by $T\ln[(1-n(p))/n(p)]$
we arrive, as before, at the system of two equations
\begin{eqnarray}
\ln {1-n(p,T)\over n(p,T)}=\int f(p,p_1) \nu(p_1,T) d\upsilon_1,
\quad p\leq p_f, \nonumber \\
-\ln \nu(p,T)=z_*(p)+\!\!\int\!\! f(p,p_1) \nu(p_1,T) d\upsilon_1,
\; p\geq p_f,
\label{systfc}
\end{eqnarray}
where $z_*(p)=\epsilon_*(p)/T$, with
\beq
\epsilon_*(p)={p^2\over 2M}+\int f(p,p_1) n_*(p_1) d\upsilon_1-\mu,
\; p\geq p_f. \quad
\eeq
In deriving the second of Eqs.~(\ref{systfc}), the logarithmic term has
been simplified by the replacement $1-\nu(p)\to 1$, noting that
the quantity $\nu(p)$ is small everywhere in momentum space.  The behavior
of $\nu(p)$ is readily explicated in two regions of momentum space
on either side of the boundary point $p_f$.  In the FC region,
upon neglecting $\nu$ on the left-hand side of the first of
Eqs.~(\ref{systfc}), this equation becomes linear, and upon
introducing the low-$T$ expansion, we have
\beq
\nu(p, T)= \alpha_0+\alpha_1T+\dots \ , \quad p\in {\cal C}.
\label{nut}
\eeq
The coefficients $\alpha_0\sim O(1),\alpha_1,\dots$
 may be evaluated
in closed form.  This expansion begins to fail near the boundary
 point $p_f$, where the ratio $f n_*(p)/T$ becomes less than unity and
the logarithmic structure of the left-hand members of Eq.~(\ref{systfc})
exhibits its influence in the emergence of
 an additional factor $\ln(1/T)$ in $\nu(T)$.  Outside the FC
region, the solution
\beq
\nu(p,T)\simeq e^{-\epsilon_*(p)/T}
\label{bol}
\eeq
is easily found, provided $\epsilon_*(p)/T\gg 1$.  Near the
boundary point $p_f$ where $\epsilon_*(p)/T\simeq 1$,
the simple formula (\ref{bol}) fails, and the quantity $\nu(T)$ again experiences
a moderate enhancement, which can be evaluated only numerically.

The impact of such enhancement on the thermodynamic quantities
of a system hosting a FC is not significant, since the range of the
domain where the logarithmic enhancement of $\nu(T)$ takes place
is too small.  As an example, let us evaluate the specific heat.
We recall that at the QCP, the Sommerfeld ratio
\beq
\gamma(T)=\int { \epsilon\over T}{\partial n(\epsilon)
\over \partial T}{dp\over d\epsilon} d\epsilon
\label{fcct}
\eeq
{\it diverges} at $T\to 0$ due to the divergence of the density
of states, which is proportional to the effective mass $M^*(T)$.
In systems with a FC, this divergence is seen to disappear because
the major term $n_*(p)$ of the sum (\ref{dec}), being $T$-independent,
does not contribute to $\gamma(T)$.  With this result, upon
inserting Eq.~(\ref{nut}) into Eq.~(\ref{fcct}), the contribution
of the FC region to $\gamma(T)$ is evaluated in a straightforward
matter, retaining the first term $\alpha_0\simeq 1$ of the expansion
(\ref{nut}).  Since, as always in thermodynamic calculations,
the condition $|\epsilon|/T\simeq 1$ applies, the result so
obtained turns out to be $T$-independent just as in standard FL
theory--- a characteristic feature of systems with a FC,
first disclosed in Ref.~\onlinecite{noz}.

Let us us now estimate the contribution to the Sommerfeld ratio
$\gamma$ from an integration region in which the function $\nu$ is
logarithmically enhanced. Since in this region, the limits of
integration are restricted by the same condition
$|\epsilon|\simeq |\epsilon_*|\leq T$, we surmise that
the magnitude of this contribution is determined by the product
of the factor $T$ and the average value of the factor
$dp/d\epsilon_*(p)$.  Analysis shows that the latter is
indeed enhanced in the boundary region; one has $dp/d\epsilon_*(p)
\propto T^{-\alpha}$, but with $\alpha<1$.  Collecting
all factors, we find that at $T\to 0$, the contribution to
$\gamma$ of the regions adjacent to the boundary point $p_f$
is suppressed as $T^{1-\alpha}\ln (1/T)$.  Accordingly, we
conclude that the emergence of a FC in significant proportion
produces a great suppression of the QCP behavior of the
Sommerfeld ratio.

\subsection{Fate of the Landau-Luttinger theorem in systems with a FC}

Here we demonstrate that the Landau-Luttinger theorem (\ref{ll})
breaks down in systems hosting a FC. It will be instructive to
pinpoint the flaw that emerges in pursuit of the standard
Luttinger argument.\cite{lut}  The analysis begins with
expression of the particle number $N$ as the integral \cite{trio}
\beq
{N\over V}=-2\int\!\!\int\limits_{-\infty}^0{\partial\over
\partial\varepsilon}\ln \left[{G(p,\varepsilon)\over
G^*(p,\varepsilon)}\right]{d{\bf p}\,d\varepsilon\over (2\pi)^4i}.
\eeq
In conventional Fermi liquids, this integral is proportional to
the difference $\ln\left(G(p,\varepsilon=0)/G^*(p,\varepsilon=0)\right)
-\ln\left(G(p,\varepsilon=-\infty)/ G^*(p,\varepsilon=-\infty)\right)$.
In turn, the value of the ratio $\ln\left(G(p,\varepsilon)/
G^*(p,\varepsilon)\right)$ is proportional to the argument
$\varphi(\varepsilon)$ of the complex function $\ln G(p,\varepsilon)$.
One has \cite{trio} $\varphi(-\infty)=\pi$ independently of the
presence or absence of a FC in the system.  Since
${\rm Im}\, G^{-1}(p,\varepsilon)=0$, a nonzero value of the integral
in momentum space is obtained by integrating over a
domain where ${\rm Re}\, G(p,\varepsilon=0)>0$, yielding the customary
result $p^3_F/3\pi^2$.  However, in the FC region, the function
${\rm Re}\, G^{-1}(p,\varepsilon=0)$ vanishes identically, and for
states belonging to the FC, the result, contained in the expression
\beq
{N\over V}={p^3_i\over 3\pi^2}-2\int\limits_{p_i}^{p_f}\!
\int\limits_{-\infty}^0{\partial\over\partial\varepsilon}
\ln\left[{G(p,\varepsilon)\over G^*(p,\varepsilon)}\right]
{d{\bf p}\,d\varepsilon\over (2\pi)^4i},
\label{nv}
\eeq
becomes uncertain. It is treated by dividing the region of energy
integration into two intervals, namely from $-\infty$ to a small
negative value $\delta$, and from $\delta$ to 0.  According to
Eq.~(\ref{green}), $G(p,\varepsilon=\delta)$ is simply $1/\epsilon$,
in agreement with the property (\ref{volg}).  Then the integral over
the first of the two intervals vanishes, since, at
$\varepsilon=\delta$, the argument of the ratio
$\ln\left(G(p,\varepsilon=\delta)/ G^*(p,\varepsilon=\delta)\right)$
coincides with that at $\varepsilon=-\infty$.  As a result, we are
left with the integral over the second interval, containing
$\ln\left(G(p,\varepsilon=0)/ G^*(p,\varepsilon=0)\right)$,
which can be evaluated by going around the pole of the Green
function (\ref{green}).  As a result, we are led to the Landau
formula (\ref{part}), which now reads explicitly
\beq
{N\over V}= {p^3_i\over 3\pi^2}+2\int\limits_{p_i}^{p_f}
n_*(p) {d{\bf p}\over (2\pi)^3}.
\label{lanp}
\eeq
Thus we conclude that in systems having a FC, the Landau-Luttinger
theorem (\ref{ll}) is violated, yet the Landau postulate
(\ref{part}) remains intact.

\section{Merging of single-particle levels in finite Fermi systems}

A phenomenon analogous to swelling of the Fermi surface in infinite
Fermi systems also exists in finite systems.  The systems implicated
include atomic nuclei, whose quantitative description within
the FL framework was elaborated by A.\ B.\ Migdal.\cite{migt}
In their conventional wisdom, textbooks teach us that under
variation of input parameters, two neighboring single-particle
levels may repel or cross one other.  However, this familiar
dichotomy overlooks a third alternative: levels can in fact
merge.\cite{haochen}

This phenomenon, missing in the theory of finite Fermi systems,
is made possible by the variation of single-particle energies
with level occupation numbers---a property central to Landau
theory.  The primary condition for merging to occur is that the
Landau-Migdal interaction function $f$ is repulsive in coordinate
space, which holds for the effective $nn$ and $pp$ interactions
in the nuclear interior \cite{migt} and for the
electron-electron interaction in atoms.

\subsection{Schematic model of merging of single-particle levels}

Following Ref.~\onlinecite{haochen}, consider a schematic model
involving three equidistant neutron levels, separated by an energy
distance $D$ in an open shell of a spherical nucleus.  The levels
are denoted $-$, $0$, and $+$, in order of increasing energy. The
single-particle energies $\epsilon_{\lambda}$ and wave functions
$\varphi_{\lambda}({\bf r})=R_{nl}(r)\Phi_{jlm}({\bf n})$ are solutions
of
\begin{equation}
[p^2/2M+\Sigma({\bf r},{\bf p})]\varphi_{\lambda}({\bf r}) =
\epsilon_{\lambda}\varphi_{\lambda}({\bf r}),
\end{equation}
where $\Sigma$ denotes the self-energy. In even-even spherical
nuclei, which in their ground states have total angular momentum
$J=0$ due to pairing correlations, the energies $\epsilon_{\lambda}$
are independent of the magnetic quantum number $m$ associated
with the total single-particle angular momentum $j$. We suppose
that the level $-$ is filled, the level $+$ is empty, and $N$
neutrons are added to the level $0$, changing the density
$\rho(r)$ by $\delta\rho(r)=NR^2_{n_0l_0}(r)/4\pi$.

In what follows we shall retain only a major, spin- and
momentum-independent part $V$ of the self-energy $\Sigma$ and a
dominant, $\delta(r)$-like portion of the Landau-Migdal interaction
function $f$.  Accordingly, the FL relation between $\Sigma $
and $\rho$ responsible for the variation of
$\epsilon_{\lambda}(n)$ with $n$ reduces to \cite{migt}
\begin{equation}
\delta V(r)=f[\rho(r)]\delta \rho(r).
\label{relation1}
\end{equation}

When particles are added to the system, all energy levels are shifted
somewhat, but the level that receives the particles is affected more
strongly than the others. For the sake of simplicity, the diagonal and
nondiagonal matrix elements of $f$ are assigned the respective values
\begin{eqnarray}
u&=&\int R_{nl}^2(r)f\left[\rho(r)\right]R^2_{nl}(r)r^2dr/4\pi, \nonumber\\
w&=& \int R_{nl}^2(r)f\left[\rho(r)\right]R^2_{n_1l_1}(r)r^2dr/ 4\pi,
\label{mel}
\end{eqnarray}
independently of the quantum numbers $nl,\,n_1l_1$.

Based on these assumptions and results, the dimensionless shifts
$\xi_k(N)=\left[\epsilon_k(N)-\epsilon_k(0)\right]/D$ for
$k=0,+,-$ are given by
\begin{equation}
  \xi_0(N)=n_0U, \quad \xi_+(N)=\xi_-(N)=n_0W,
\label{en1}
\end{equation}
where $n_k=N_k/(2j_k+1)$ is the occupation number of the level $k$,
$U=u(2j_0+1)/D$, and $W=w(2j_0+1)/D$. It is readily verified that
if $fp_FM/\pi^2\sim 1$, where $p_F=\sqrt{2M\epsilon_F}$ and
$\epsilon_F$ is the Fermi energy, then the first of the
integrals (\ref{mel}) has a value $u\simeq \epsilon_F/A$
and therefore $U\sim 1$, since $D\sim \epsilon_F/A^{2/3}$
in spherical nuclei.

According to Eqs.~(\ref{en1}) at $(U-W)>1$, the difference
$d(N)=1+\xi_+(N)-\xi_0(N)$ changes sign at $n_{0c}=1/(U-W)$,
before filling of the level $+$ is complete.  At $n_0>n_{0c}$, in
the standard scenario provided by Hartree-Fock theory, all added
quasiparticles must resettle into the empty single-particle
level $+$. However, not all of the quasiparticles can take part
in the migration process, since the situation would then be
reversed, and the roles of the levels interchanged: the formerly
empty level, lying above the formerly occupied one, would have
the maximum positive energy shift, rendering migration impossible.
Thus, the standard Fermi-liquid filling scenario, which prescribes
that {\it one and only one} single-particle level lying exactly
at the Fermi surface can remain unfilled, while all others must
be completely occupied or empty, encounters a contradiction.

This contradiction is resolved as follows.\cite{haochen} Migration
occurs until the single-particle energies of the two levels in
play coincide.  As a result, both of the levels, $0$ and $+$,
become {\it partially} occupied---an impossible situation for
the standard Landau state. Solution of the problem reduces to
finding the minimum of the relevant energy functional
\begin{equation}
E_0=\epsilon_0(0)N_0+\epsilon_+(0)N_++{1\over 2}\left[u(N^2_0+N^2_+)
+2wN_0N_+\right]
\label{energy}
\end{equation}
with $N_k=\sum_m n_{km}$, through a variational condition
\begin{equation}
{\delta E_0\over\delta n_{0m}}={\delta E\over\delta n_{+m_1}}=\mu,
\qquad
\forall m, m_1,
\label{varn}
\end{equation}
where $\mu$ is the chemical potential.  Eqs.~(\ref{varn}) are
conveniently rewritten as conditions
\begin{eqnarray}
\epsilon_0(N)&=&\epsilon_0(0)+N_0u+N_+w=\mu, \nonumber\\
\epsilon_+(N)&=&\epsilon_+(0)+N_0w+N_+u=\mu
\label{en3}
\end{eqnarray}
for coincidence of the single-particle energies $\epsilon_0$
and $\epsilon_+$, which, at $N>N_c=(2j_0+1)/(U-W)$, yield
$N_0={1\over 2}(N+N_c)$ and $N_+={1\over 2}(N-N_c)$.

\begin{figure}[t]
\includegraphics[width=0.4\textwidth,height=0.4\textwidth]{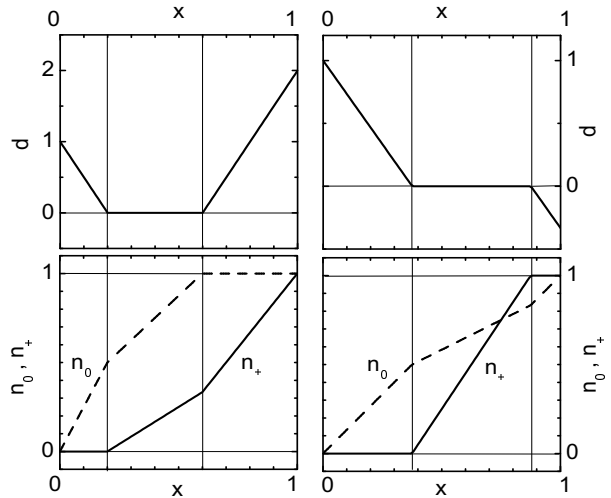}
\caption{Top panels: Dimensionless distance $d=(\epsilon_+-\epsilon_0)/D$
between levels $+$ and $0$ as a function and $0$ as a function of the ratio $x=N/(2j_0+2j_++2)$. Lower panels: Occupation numbers $n_k$ for levels $0$ and $+$. Input parameters: $U=V=3, W=1$. For the left column, the ratio $r\equiv (2j_0+1)/(2j_++1)=2/3$; for the right, $r=3$.}
\label{fig:two}
\end{figure}

Results from numerical calculations are plotted in Fig.~\ref{fig:two},
which consists of two columns, each made up of three plots. The
upper panels show the dimensionless ratio
$d(x)=\left]\epsilon_+(x)-\epsilon_0(x)\right]/D$ versus
$x=N/(2j_0+2j_++2) \in [0,1]$. The lower panels give the occupation
numbers $n_+(x)$ and $n_0(x)$. We observe that there are three
different regimes: in two of them $d\neq 0$ and there exist
well-defined single-particle excitations, and in the third, the
energies of the levels 0 and + coincide at zero. Passage through
the three regimes can be regarded as a second-order phase transition,
with the occupation number $n_+$ treated as an order parameter.

The single-particle levels remain merged until one of them is
completely filled. If the level $0$ fills first, as in the left
column of Fig.~\ref{fig:two}, then under further increase of $N$,
quasiparticles fill the level $+$, signaling that the distance
$d(N)$ again becomes positive. This behavior resembles the
repulsion of two levels of the {\it same symmetry} in quantum
mechanics, although here one deals with the single-particle levels of
{\it different symmetry}.  In a case where level $+$ becomes fully
occupied before the level 0 does, as in the right column,
the distance $d(N)$ becomes negative, and the two levels just
cross each other at this point.  Interestingly, analogous behavior
has been discovered recently \cite{vol2010a} in numerical
calculations of single-particle spectra of multilayered
systems.

\subsection{Competition between shell-model effects and merging of
single-particle levels}

As is well known that the stability of atomic nuclei hinges
critically upon negative shell corrections to the Weizs\"acker
formula for ground-state energies. These corrections are maximal
in magic nuclei such as Sn$^{132}$ and Pb$^{208}$, where the density
of single-particle levels near the Fermi surface drops sharply due
to the existence of  an
 energy gap between the last occupied and
first unoccupied levels.  Such behavior is in complete agreement
with the idea of an atomic nucleus ``as a gas of interacting
quasiparticles''.\cite{migt} The magnitude of shell effects
decreases with increasing atomic numbers, primarily because the
magnitude of the magic gap falls off rapidly in heavy and superheavy
nuclei. With merging phenomenon coming into play, the situation
changes. For one thing, merging ensures a negative contribution
to the ground-state energy, despite the fact that the density
of the single-particle levels exhibits an opposing trend, reaching
a maximum at the Fermi surface, rather than a minimum, as in the
conventional shell model.

In effect, the existence of gaps in
filling prevents the merging phenomenon from working in full force.
However, in heavy and superheavy atomic nuclei where this gap
becomes smaller and smaller, the FC correction to the Weizs\"acker
ground state-energy grows in importance, creating the opportunity
for level merging to be an important playmaker in the formation
of new stability islands.

\subsection{Merging of single-particle levels in an external magnetic field}

In this subsection, we employ the simplest model of fermion condensation,
proposed in Ref.~\onlinecite{noz}, to investigate the impact of an
external magnetic field on the FC spectrum and demonstrate that
the presence of this field does not prevent the Fermi surface
from swelling.  The analysis is restricted to the simplest form
of the FL energy functional, namely
\beq
E={1\over 2M}\int \tau({\bf r})d{\bf r} + {1\over 2}\int\!\!\!\int
f({\bf r}-{\bf s})\rho^2({\bf r},{\bf s})d{\bf r}d{\bf s},
\label{fle}
\eeq
in which $f$ stands for the Landau interaction function. The
density matrix $\rho$ and kinetic-energy density $\tau$ are
respectively given by
\beq
\rho({\bf r},{\bf s})=\sum n_i \Psi_i^*({\bf r})\Psi_i({\bf s}), \;
\tau({\bf r})= \sum n_i\nabla\Psi_i^*({\bf r})\nabla\Psi_i({\bf r}),
\label{dens}
\eeq
in terms of the true quasiparticle occupation numbers $n_i$ of the single-particle states. Imposition of the external magnetic field ${\bf H}$ leads to a set of coupled equations
\beq
{({\bf p}-e{\bf A}({\bf r}))^2\over 2M}\Psi_j({\bf r})
+\int f({\bf r}-{\bf s})\rho({\bf r},{\bf s})\Psi_j({\bf s})d{\bf s}=
E_j\Psi_j({\bf r}),
\label{wf}
\eeq
for the eigenvalues $E_j$ and eigenfunctions $\Psi_j$,
where ${\bf A}$ is the corresponding vector potential.  Specifically,
we address 2D homogeneous electron systems, employing the Landau gauge
in which $A_x=-Hy$ and $A_y=0$. Generally, each of the equations
(\ref{wf}) contains a sum of an infinite number of terms involving
different wave functions $\Psi_i$. This complication is
normally surmounted by working within a quasiclassical
approximation,\cite{trio} such that
\beq
\rho({\bf r},{\bf s},{\bf A})= e^{iA({\bf r})({\bf r}-{\bf s})}
\rho_0({\bf r}-{\bf s}),
\label{qa}
\eeq
where $\rho_0$ is the well-known quasiparticle density matrix
for homogeneous matter. However, the quasiclassical approximation fails
in systems having long-range interaction functions of the kind
responsible for flattening of the single-particle spectrum
$\epsilon(p)$ beyond the QCP.

On the other hand, it is just the long-range character of the relevant
effective interactions that allows one to proceed while avoiding
the approximation (\ref{qa}).  Indeed, suppose the correlation radius
$\xi$ specifying the behavior of $f$ at large distances is in
excess of the Larmor radius $r_H=1/\sqrt{eH}$.  In the sum over
$i$ implicit in Eq.~(\ref{wf}), the diagonal term is dominant,
and the initial structure $\Psi(x,y)=e^{ip_x x}\chi(y)$ of the
corresponding wave functions is recovered.  Neglecting corrections
of order $r^2_H/\xi^2$ coming from nondiagonal contributions,
we are then left with the Landau-like equation
\beq
{1\over 2M}\left(p^2_y+e^2H^2(y-y_0)^2\right)\chi_j(y)=(E_j-fn_j)\chi_j(y),
\label{modlan}
\eeq
where $f$ is the effective coupling constant.  As usual, $y_0$
specifies the location of the Larmor circle, while the occupation
number $n_j=[1+\exp(\epsilon_j/T)]^{-1}$ is itself
dependent upon the deviation $\epsilon_j=E_j-\mu$ of the single-particle
level from the chemical potential $\mu$.  Now, according to
Eq.~(\ref{modlan}), the initial equidistant spectrum
$E^0_j\equiv E_j(H=0)=\omega^0_c(j+1/2)$, with $\omega^0_c=eH/M$,
gives way to a modified level scheme determined by integer
solutions of nonlinear algebraic equation
\beq
\epsilon(x)=\omega_c^0x-\mu+fn(x).
\label{spech}
\eeq
Recalling that $n(\epsilon=0)=1/2$, Eq.~(\ref{spech}) may be
recast in the form
\beq
n(\epsilon)={\epsilon\over f}+{1\over 2}\left(1-{y\over r}\right),
\label{nfc}
\eeq
where $y=x-(\mu-f/2)/\omega_c^0$ gives the displacement of the
single-particle levels from the Fermi surface.  Eq.~(\ref{nfc})
is reminiscent of the corresponding equation for the NFL occupation
numbers in the Nozi\`eres model,\cite{noz} developed for homogeneous
matter with an interaction function $f({\bf r}-{\bf s})={\rm const.}$
of infinite radius.  The QCP of the model \cite{noz} is located
at $f=0$, with the flattening phenomenon in effect at $f>0$.

\begin{figure}[t]
\includegraphics[width=0.8\linewidth,height=0.5\linewidth]{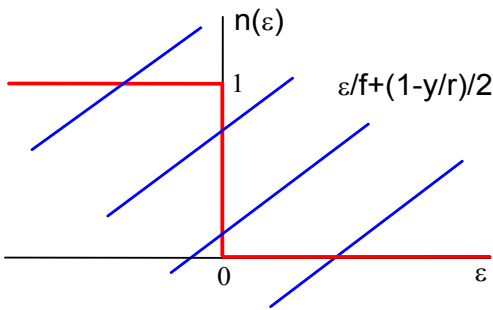}
\caption{Graphical illustration of the solution of Eq.~(\ref{nfc})}
\label{fig:noz_sol}
\end{figure}

Eq.~(\ref{nfc}) can easily be solved graphically at $T=0$
(see Fig.~\ref{fig:noz_sol}).  This is done by plotting
the kink $n(\epsilon)$, together with the set of straight lines
representing the right side of Eq.~(\ref{nfc}), against the input
parameter $y$.  At $y>r$, the right side of Eq.~(\ref{nfc})
has the form $\epsilon-a$, with $a>0$.  In this case, crossing
points lie in the right half-plane $\epsilon>0$ where $n=0$,
implying that the initial Landau spectrum $E^0_j=\omega_c(j+1/2)$
is unaffected by the interactions.  On the other hand, at
$y<-r$, the crossing points transfer to the left half-plane
$\epsilon<0$ where $n=1$, so that the equidistant Landau
spectrum is simply shifted by the constant $f$.

What happens in the interval $-r<y<r$?  In this case, as seen
from Fig.~\ref{fig:noz_sol}, the crossing points land on the
vertical segment of the kink, yielding the solution $\epsilon=0$;
Eq.~(\ref{nfc}) reduces to relation
\beq
n_*(y)={1\over 2}\left(1-{y\over r}\right),\quad -r<y<r,
\label{nnoz}
\eeq
for finding the NFL occupation numbers $n_*(y,H)$ at $H\neq 0$.
Comparison with the corresponding result $n_*(p)=(\mu-p^2/(2M))/f$
for the Nozi\`eres model \cite{noz} shows that the swelling of the
Fermi surface inherent in this model persists, the range of the
FC domain being unaffected by imposition of the magnetic field.  At
finite $T$, analysis of Eq.~(\ref{nfc}) confirms that in the FC
domain, the derivative $d\epsilon(x)/dx$ becomes proportional
to $T$, again in accordance with the results obtained in homogeneous
matter at $H=0$.\cite{noz}

\section{Lifshitz phase diagram: Evolution from the Landau state to
the FC state}

Low-temperature phase transitions occurring in Fermi systems have
already been studied for a hundred years.  As a rule, these are
second-order phase transitions, epitomized in the phenomenon
of superconductivity discovered by Kamerlingh-Onnes in 1911.  By
contrast, experimental studies of topological phase transitions
(TPTs) have gained prominence only during the last decade.  This
disparity stems from the fact that low-temperature, second-order phase
transitions associated with violation of one or another of the
Pomeranchuk stability conditions may arise in weakly or moderately
correlated systems, as in the example of superconductivity.  In
the case of TPTs, it is instead the necessary stability condition
(\ref{necs0}) for the Landau state that suffers a breakdown, an
eventuality that can be realized only in strongly correlated Fermi
liquids, notably in dense films of liquid $^3$He and the dilute
2D electron liquid in MOSFETs.  Nothing was known about such systems
prior to recent experimental and technological advances. To our
knowledge, the possibility of a many-fermion ground state with
a multi-connected Fermi surface was first entertained by Fr\"olich in
1950.\cite{frolich} Theoretical consideration of topological
aspects of the electron liquid in solids dates back to an article
by I.\ M.\ Lifshitz published ten years later.\cite{lifshitz}

The Lifshitz analysis requires specific information on the structure
of the single-particle spectrum $\epsilon({\bf p})$, which is often not
available because this spectrum may be altered dramatically upon
switching on the effective interactions between quasiparticles.\cite{ks}
For many years, insights into the effects of these interactions
were pursued within
the Hartree-Fock framework.\cite{Vary,Aguilera,baym,schofield} It was
not until 1990 that such limited descriptions began to be replaced
by more powerful FL approaches.\cite{ks,vol,noz,physrep,zb,shagp,zkb,
khodel2007,shagrev,prb2008,jetplett2009,prb2010}

In dealing with homogeneous systems, any TPT is associated with a
change in the number of roots of equation \cite{volrev}
\beq
\epsilon(p;n,\rho)=0,
\label{topeq}
\eeq
where $\epsilon(p;n)$ is the spectrum of single-particle excitations
measured from the Fermi surface and evaluated self-consistently
with the $T=0$ ground-state quasiparticle momentum distribution
$n(p)$.  In conventional homogeneous Fermi liquids such as 3D
liquid $^3$He, Eq.~(\ref{topeq}) has a single root, the Fermi
momentum $p_F$; hence the corresponding ground-state quasiparticle
momentum distribution is simply
  $n_F(p)=\theta(p_F-p)$.

The QCP corresponds to the particular case of a topological
transition for which the bifurcation point $p_b$ coincides
with the Fermi momentum $p_F$. By its definition, at the QCP
the effective mass diverges and therefore the Fermi velocity
$v_F(\rho_{\infty})=\left(d\epsilon(p,\rho_{\infty})/dp\right)_0$
vanishes. In this situation, the spectrum is constrained to have
an inflection point,\cite{prb2005} i.e.\ two additional roots
of Eq.~(\ref{topeq}) emerge simultaneously at the Fermi surface,
implying that $\epsilon(p)\propto (p-p_F)^3$ at the QCP itself.
It will be seen that such a situation is not typical.  As a matter
of fact, as the system moves into the quantum critical regime from
the FL domain, Eq.~(\ref{topeq}) must {\it already} acquire two
new roots $p_1,p_2\neq p_F$ at some critical value $\lambda_{\qcp}$
of the input parameter $\lambda$.  Accordingly, the Fermi surface
becomes multi-connected, but with the ground-state momentum distribution
  $n(p)=\theta(p_1-p)-\theta(p_2-p)+\theta(p_F-p)$,
 still satisfying the usual FL relation $n^2(p)=n(p)$. On the other hand, those
solutions of Eq.~(\ref{lansp}) emerging at some critical coupling
$\lambda_{\fc}$ and featuring a completely flat portion of the spectrum
$\epsilon(p)$ correspond to solutions of Eq.~(\ref{topeq}) having an
{\it infinite} number of roots, or equivalently, solutions with an
infinite number of sheets of the Fermi surface. We reiterate that
in the associated FC phase, the fundamental relation $n^2(p)=n(p)$
of standard FL theory no longer holds; {\it partial occupation}
of single-particle states must occur.

Numerous calculations have demonstrated that as a rule, the
ratio $\lambda_{\fc}/\lambda_{\qcp}$ lies between 1 and 2.  When
the coupling constant $\lambda$ increases beyond $\lambda_{\qcp}$
into the regime of instability of the Landau state, a striking
sequence of rearrangements of the single-particle degrees of
freedom is witnessed,\cite{zb} as pockets of the Fermi surface
breed and proliferate rapidly, in resemblance to a fractal cascade.
Our first task in the present section is to examine the details of
the evolution of the topology of the ground state in the
Lifshitz phase diagram over the coupling interval
$\lambda_{\qcp}<\lambda<\lambda_{\fc}$. Our second goal is
to expose the intricate evolution, with increasing temperature, of
the single-particle spectrum $\epsilon(p)$ in phases with
having a multi-connected Fermi surface.

\subsection{Generic features of topological transitions in Fermi
liquids}

In principle, the bifurcation $p_b$ of Eq.~(\ref{topeq})
giving rise to the first TPT can emerge at any point of momentum
space.  Leaving aside the case addressed in the previous section,
in which the transition leads to a swollen Fermi surface, we are
left with the following options.  The first and common option
is $p_b=0$. The second and particular one is the QCP for which
$p_b=p_F$.  In the third case $0<p_b<p_F$, the bifurcation point
$p_b$ is a local maximum of the spectrum $\epsilon(p)$, since
one has $\epsilon(p)\leq 0$ throughout the Fermi volume. This
implies that between $p_b$ and $p_F$ there also exists a local
minimum of $\epsilon(p) $, because by definition the spectrum
$\epsilon(p) $ changes sign at $p=p_F$.  Thus, if the bifurcation
point is located between 0 and $p_F$, the equation $d\epsilon(p)/dp=0$
has {\it at least} two roots in this momentum interval. Simple
analysis demonstrates that existence of at least two roots of
this equation is also inherent in the fourth and last case,
in which Eq.~(\ref{topeq}) bifurcates at a momentum greater than
$p_F$ and the bifurcation point is a local minimum of the
curve $\epsilon(p)$.

A prerequisite for the occurrence of the bifurcations in Eq.~(\ref{topeq})
is a marked momentum dependence of the interaction function
$f({\bf p},{\bf p}_1)$. Such a momentum dependence is often associated
with forces of long range in in coordinate space that are induced in
advance of an impending second-order phase transition.  As a pertinent
illustration, let us consider the situation in the vicinity of pion
condensation, a phase transition in dense neutron matter predicted
by A.\ B.\ Migdal.\cite{migrevmod,migphysrep}  We restrict ourselves
to the case of moderate pion fluctuations where the Ornstein-Zernike
approximation, allowing convenient treatment of the fluctuation
contribution to the interaction function $f$, is still applicable.
In this case, the exchange contribution $f_{\pi\pi}$ to the interaction
function $f$ has the form \cite{migrevmod}
\beq
f_{\pi\pi}(q)= {g \over \kappa^2(\rho)+(q^2/q^2_c-1)^2},
\label{tio}
\eeq
with coupling constant $g$ and the critical momentum taken as
  $q_c\simeq (0.7\div 1.0)\,p_F$.
  The stiffness coefficient $\kappa^2(\rho)$
vanishes at the critical density $\rho_c$, specifying the point
where pion condensation sets in. For simplicity, we choose
  $g=1/(2m^2_{\pi})$,
 corresponding to bare $\pi NN$ vertices,
and study the behavior of quantities versus $\kappa(\rho)$.

In the Landau state, corrections to the neutron spectrum
$\epsilon(p)$ associated with the exchange of pion fluctuations
are evaluated in closed form by means of Eq.~(\ref{lansp}),
based on the quasiparticle distribution $n_F(p)$.  Substituting
the expression (\ref{tio}) and integrating, we obtain
\beq
{d\epsilon(p)\over dp}={p\over M}+{gq^4_c\over 16\pi^2 p^2}
\left( -{1\over 2}L(p)+ \frac{p^2{+}p_F^2{-}q_c^2}{\kappa q_c^2}
A(p)\right),
\label{deriv}
\eeq
where
\beq
L(p)=\ln \frac{[(p{+}p_F)^2{-}q^2_c]^2+\kappa^2 q_c^4}
{[(p{-}p_F)^2{-}q^2_c]^2+\kappa^2 q_c^4}
\eeq
and
\beq
A(p)= \arctan\frac{(p_F{+}p)^2{-}q^2_c}{\kappa q^2_c}
-\arctan\frac{(p_F{-}p)^2-q^2_c}{\kappa q^2_c}.
\eeq
Further integration can be performed analytically,\cite{zkb}
but the results are too cumbersome to present here.

\begin{figure}[t]
\includegraphics[width=0.8\linewidth,height=1.\linewidth]{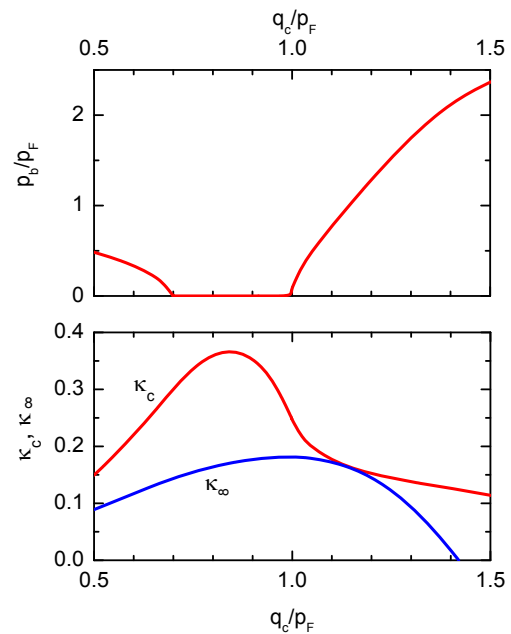}
\caption{Upper panel: Position of the bifurcation point $p_b$ in units of the Fermi momentum, versus the critical wave number $q_c$ (also in units of $p_F$). Lower panel: critical parameters $\kappa_b$ and $\kappa_{\infty}$ as functions of $q_c/p_F$.}
\label{fig:nm_bifur}
\end{figure}

Results of numerical calculations displayed in the upper panel of
Fig.~\ref{fig:nm_bifur} demonstrate that a new root of Eq.~(\ref{topeq}),
lying at the origin, appears already at $\kappa=\kappa_b\simeq 0.4$,
signaling that the Landau state has become unstable {\it well before}
the system attains the point of pion condensation. At customary
values of the critical momentum $q_c$, one has $p_b=0$. However,
as $q_c$ increases to greater values, $p_b$ rapidly moves toward the
Fermi momentum and crosses the Fermi surface at $q_c\simeq 1.14\,p_F$.
The dependence of the critical parameter $\kappa_b$ on the wave
number $q_c$, depicted in the lower panel, shows that the largest
values of $\kappa_b$ are achieved just in the preferred range
$q_c/p_F\sim 0.7-1.0$. The value $\kappa_{\infty}$ of $\kappa$
at which the border of the instability region reaches the Fermi
momentum $p_F$ is also plotted in the lower panel of
Fig.~\ref{fig:nm_bifur}. The resulting curve lies below the
curve of $\kappa_b(q_c)$ everywhere except for the point of
contact at $q_c\simeq 1.14\,p_F$.

\begin{figure}[t]
\includegraphics[width=1.\linewidth,height=0.9\linewidth]{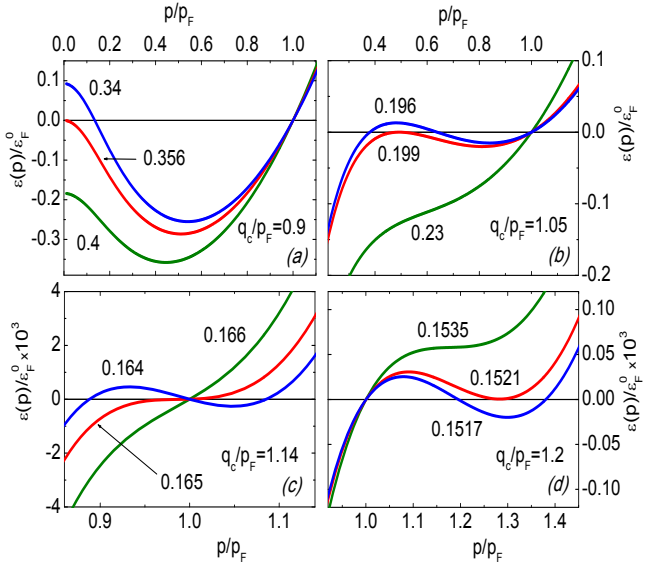}
\caption{Neutron quasiparticle spectra $\epsilon(p)$ (in units of
$\varepsilon^0_F$) evaluated for $q_c=0.9\, p_F$ (panel (a)),
$q_c=1.05\, p_F$ (panel (b)), $q_c=1.14\, p_F$ (panel (c)), and
$q_c=1.2\, p_F$ (panel (d)).  Corresponding values of the
parameter $\kappa$ are indicated near the curves.}
\label{fig:nm_spectra}
\end{figure}

The corresponding evolution of the neutron spectra is illustrated
by panels (b)--(d) of Fig.~\ref{fig:nm_spectra} where the spectra
$\epsilon(p)$ calculated for $q_c=1.05\,p_F, 1.14\,p_F$, and
$1.2\,p_F$ are drawn. We see that at $q_c< p_F$ the bifurcation
point $p_b$ does emerge quite close to the origin. This topological
transition plays important role in the acceleration of neutrino
cooling of neutron stars,\cite{vkzc} since the presence of a
new pocket of the Fermi surface in the low-momentum region
lifts the ban on neutron $\beta$-decay and activates the highly
efficient direct-Urca process.

As seen from panels (d) and (e) of Fig.~\ref{fig:nm_spectra},
beyond the Fermi liquid QCP the curve $\epsilon(p)$ crosses
the Fermi level three times, the $T=0$ Landau momentum distribution
$n_F(p)=\theta(p_F-p)$ is rearranged, and the Fermi surface
becomes multi-connected. In this instance, where there is a single
bubble in the filling pattern, the new $T=0$ occupation numbers
$n(p,T=0)$ are given by an alternating sequence of two numbers,
1 and 0.  The occupancy is $n(p)=1$ at $p<p_1$, while at $p_1<p<p_2$,
there is a gap in filling where $n(p)=0$, and at $p_2<p<p_3$, the
occupancy is once again $n(p)=1$.  This is a typical topological
phase transition, at which no symmetry inherent in the ground
state is violated.

\subsection{Prerequisites for the proximity of the first TPT and the QCP}

Interest to the mechanisms underlying violation of FL theory
has been sparked by the discovery of NFL behavior of electron
systems present in numerous heavy-fermion metals.  It is a
specific and significant feature of these NFL phenomena that
the bifurcation point $p_b$ lies close to the Fermi momentum
$p_F$. To clarify the conditions necessary for the proximity
of $p_b$ and $p_F$, it is natural to begin with the Taylor
expansion of the group velocity,
\beq
v(p)=v_F+v_1{p-p_F\over p_{\infty}}+{1\over 2}v_2{(p-p_F)^2\over p^2_F},
\label{vt}
\eeq
applicable in the immediate vicinity of the Fermi surface,
which yields the spectrum
\beq
\epsilon(p)=v_F(p-p_F)+{1\over 2p_F}v_1(p-p_F)^2 +{1\over 6p^2_F} v_2(p-p_F)^3.
\label{spt}
\eeq
Upon inserting this expression into Eq.~(\ref{topeq}), we observe
(once again) that {\it already} on the disordered side of the QCP where
$v_F(\rho)$ is still {\it positive}, this equation picks up
two additional real roots:
\beq
p_{1,2}-p_F= -p_F{3v_1\over  2v_2}\left(1\pm \sqrt{1-{8v_F(\rho)v_2
\over 3v^2_1}}\right).
\label{rootv}
\eeq
Accordingly, the first TPT occurs at a critical density $\rho_t$ where
the Fermi velocity is given by $v_F(\rho_t)=3v^2_1/8v_2$.  This
analysis is self-consistent as long as the ratio $v_1/v_2$ is
small; otherwise at least one of the roots lies too far from
the Fermi momentum for the Taylor expansion (\ref{vt}) to be
valid.  In conventional Fermi liquids, where $f({\bf p},{\bf p}_1)$
is a smooth function in momentum space and hence the coefficients
$v_1$ and $v_2$ are of the same order, the stated condition is
difficult to satisfy. However, the situation becomes more favorable in the vicinity of a second-order phase transition where the fluctuations are of long wavelength.
Assuming these fluctuations to be rather weak,
their contribution ${\cal F}^{\rm sf}$ to the interaction function $f$
can be accounted for within well-known Ornstein-Zernike (OZ) approximation to yield
\beq
{\cal F}^{\rm sf}_{\alpha\beta\gamma\delta}({\bf p},{\bf p}_1)
\to g^2 {\bm \sigma}_{\alpha\beta}{\bm \sigma}_{\gamma\delta}
\chi(|{\bf p}-{\bf p}_1|).
\label{sfl}
\eeq
Here $g$ is the spin-fluctuation vertex and $\chi(q)=4\pi/(q^2+\xi^{-2})$
is the spin susceptibility, $\xi$ being the correlation radius,
which diverges at the second-order phase transition point.
If the spin-fluctuation term is well-pronounced, then $\xi$
substantially exceeds the distance between particles.  The
applicability of the OZ approximation requires that the inequality
$(g^2/\pi v^0_F) \ln (p_F\xi)\leq 1$ should be met.\cite{jetplett2009}

The spin-fluctuation contribution $v^{\rm sf}(p)$ to the group
velocity $v(p)$ is evaluated straightforwardly, employing the identity
${\bm \sigma}_{\alpha\beta}{\bm \sigma}_{\gamma\delta}=
{3\over 2}\delta_{\alpha\delta}\delta_{\gamma\beta}-
{1\over 2}{\bm \sigma}_{\alpha\delta}{\bm \sigma}_{\gamma\beta}$ to
obtain
\beq
v^{\rm sf}(p)={3g^2\over 2\pi }\ln{1+\xi^2 (p-p_F)^2\over 4p^2_F\xi^2}.
\label{vsf}
\eeq
Then instead of Eqs.~(\ref{vt}) and (\ref{spt}) one obtains
(with $v_F^0=p_F/M$ and $v_2=\lambda_s v_F^0$)
$$
v(p)= {p_F\over M^*}+ v_1{p-p_F\over p_F}+ {3\over 2}\lambda_s v^0_F
{(p-p_F)^2\over p_F^2}, \qquad\qquad\quad
$$
\vskip -0.5 cm
\beq
\epsilon(p)={p_F\over M^*}(p-p_F)+v_1{(p-p_F)^2\over 2p_F}
 +\lambda_s v^0_F{(p-p_F)^3\over 2p_F^2},
\label{vb}
\eeq
where the effective mass $M^*$ is given by
\beq
{M\over M^*}= {M\over M^*_{\fl}}-3{g^2\over \pi v^0_F}\ln(2p_F\xi(\rho)),
\label{mefsf}
\eeq
$M^*_{\fl}$ being the FL effective mass evaluated without inclusion
of the fluctuation contribution (which drives the system toward the QCP),
while
\beq
\lambda_s=\xi^2 p_F^2{g^2\over \pi v^0_F}
\simeq {\xi^2 p_F^2\over \ln (p_F\xi)}\gg 1.
\label{lam}
\eeq
Since the condition $v_2/v_1\propto 1/\lambda_s\ll 1$ holds in
the present case, the proximity of the first TPT bifurcation point
to the Fermi momentum is established.

In addition, the density $\rho_{\infty}$ corresponding to the
topological QCP,  which signals instability of the Landau state,
is determined by condition
\beq
{M\over 3M^*_{\fl}(\rho_{\infty})}={g^2\over \pi v_F^0}
\ln (2p_F\xi(\rho_{\infty})).
\label{stc}
\eeq
In strongly correlated Fermi systems, the ratio $M/M^*_{\fl}$ is
suppressed, implying that the stability condition (\ref{stc}) is
violated {\it well before} the correlation radius $\xi(\rho)$ grows
to values sufficient to put the OZ approximation in jeopardy.

It is worth estimating the limits of the applicability of the Taylor
expansion (\ref{vb}).  As seen from Eq.~(\ref{vsf}), the expansion
breaks down in case the difference $p-p_F$ reaches values of
order of $1/\xi$.  At larger distances from the Fermi surface,
the logarithmic term in Eq.~(\ref{vsf}) asserts itself, introducing
logarithmic corrections to the specific heat and other thermodynamic
quantities (as shown explicitly below).

\subsection{Cascade of topological transitions in the QCP region}

\begin{figure}[t]
\includegraphics[width=0.72\linewidth,height=0.75\linewidth]{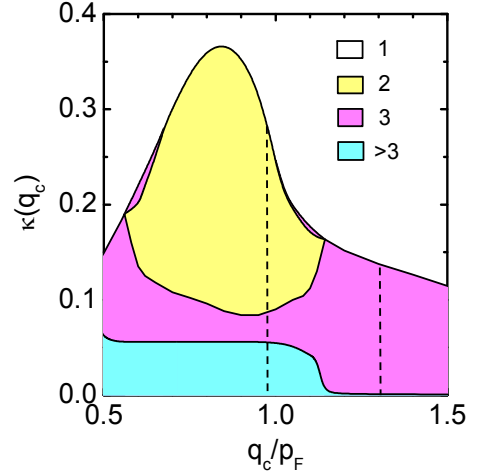}
\caption{The phase diagram of neutron matter in $(q_c,\kappa)$ variables. The Landau phase (FL) occupies the white region of the plane. The index $i$ indicates the number of sheets (or pockets) of the Fermi surface.}
\label{fig:nm_phdia}
\end{figure}

The emergence of new small pockets of the Fermi surface is an
integral feature of the QCP phenomenon, whether one deals
with 2D liquid $^3$He, high-$T_c$ superconductors, or strange
heavy-fermion metals.  A striking peculiarity of the segment
of the Lifshitz phase diagram beyond the first TPT point, revealed
in Refs.~\onlinecite{zb}, is the fast breeding of these pockets
as $\lambda$ increases beyond $\lambda_{\qcp}$. This process is
illustrated in Fig.~\ref{fig:nm_phdia}, which shows theoretical
results \cite{zkb} for the phase diagram of neutron matter in
the $(q_c,\kappa)$ plane, from calculations performed for the
interaction function (\ref{tio}) in the vicinity of the threshold
of pion condensation. The Landau state with $n(p)=n_F(p)$
occupies the white region of the diagram (labeled FL in the figure),
while the phases exhibiting new pockets of the Fermi surface populate
the shaded portion of the plane, which is separated from the FL
domain by the curve $\kappa_b(q_c)$. Ref.~\onlinecite{jpcm2004} reports
interesting results from a simple model incorporating feedback
of fermion condensation on critical fluctuations.  In particular,
the range of territory on the phase diagram in which the FC wins
the contest with the multi-pocket phase was estimated as
$0.95<q_c/p_F<1.3$ at $\kappa\simeq 0.1$.

\begin{figure}[t]
\includegraphics[width=1.\linewidth,height=0.8\linewidth]{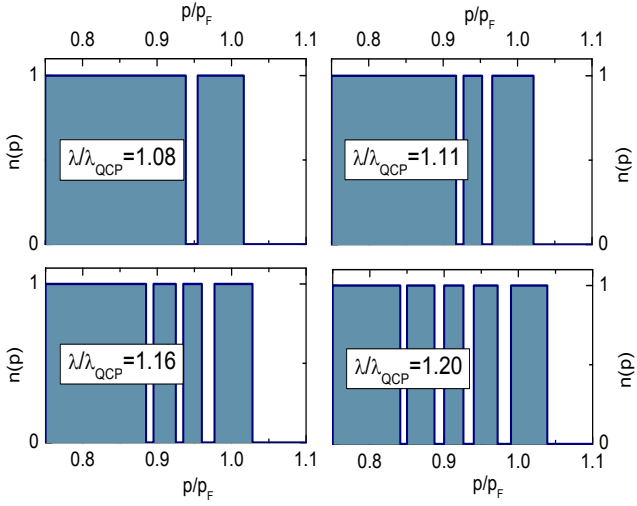}
\caption{Cascade of topological transitions, evaluated in the model with $f(q)=\lambda (q^2+\kappa^2)^{-1}$ with $\kappa=0.07\,p_F$ for four values of $\lambda/\lambda_{\qcp}=1.08$, 1.11, 1.16, 1.20.}
\label{fig:cascade}
\end{figure}

\begin{figure}[t]
\includegraphics[width=0.7\linewidth,height=0.62\linewidth]{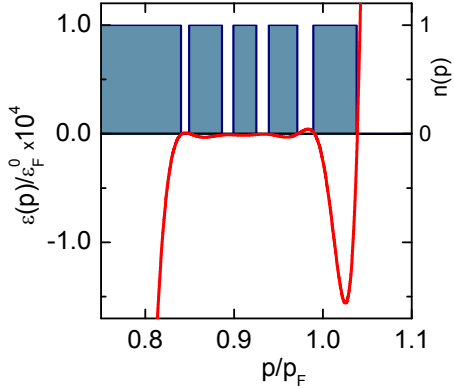}
\caption{Single-particle spectrum $\epsilon(p)$ in units of
$\epsilon_F^0=p_F^2/2M$ together with the self-consistent quasiparticle momentum distribution $n(p)$, as evaluated in Ref.~\onlinecite{zkb} within a model in which the Landau interaction function has the form $f(q)=\lambda (q^2+\kappa^2)^{-1}$, with $\kappa=0.07\,p_F$. Calculated for an effective coupling constant $\lambda$, exceeding the critical QCP value $\lambda_{\qcp}$ by 30\%, this spectrum  possesses nine nodes, corresponding to four hole pockets in momentum space.}
\label{fig:hole_pockets}
\end{figure}

As seen in Figs.~\ref{fig:cascade} and \ref{fig:hole_pockets},
the breeding phenomenon also expresses itself vigorously
in a different model \cite{zb} based on the interaction
function $f(q)=g/(q^2+\kappa^2)$.

The occurrence such cascades is associated with the condition
$n(p)\leq 1$ enforced by the Pauli principle.  As we have seen
earlier, solutions of the equilibrium equation (\ref{varfl})
always exist, but these solutions violate the restriction
$n(p)\leq 1$ until the coupling constant $\lambda$ attains
a critical value $\lambda_{\fc}$. It is this restriction that
triggers a cascade of TPTs in the segment of the Lifshitz
phase diagram defined by $\lambda_{\qcp}<\lambda<\lambda_{\fc}$.
Indeed, in the TF problem where no such restriction is present,
no topological transitions exist. Details of the transitions
between phases with a multi-connected Fermi surface and phases
hosting a FC remain unclear at this point.

In this section, we discuss results of numerical calculations
of single-particle spectra $\epsilon(p,T)$ and quasiparticle
momentum distributions $n(p,T)$ beyond the first TPT.\cite{prb2008}
As we have seen, (i) these transitions are associated with the
violation of the necessary stability condition (\ref{necs0}) for the
Landau state at $T=0$, in which $n(p)=\theta(p_F-p)$, and (ii)
there exist different alternatives for the rearrangement of
the Landau state beyond the transition point.  The first possibility
is the emergence of a small pocket of the Fermi surface; hence,
the new quasiparticle momentum distribution
$n(p,T=0)=\theta(p_1-p)-\theta(p_2-p)+\theta(p_F-p)$ features
a combination of three kinks. The second possibility amounts
to a swelling of the Fermi surface, so that the ne momentum distribution
$n(p,T=0)$ is continuous over a momentum interval $[p_1,p_2]$
embracing the boundary momentum $p_F$.  In this momentum interval,
the spectrum $\epsilon(p,T)$ is completely flat at $T=0$, and,
according to Eq.~(\ref{spte}), it is inclined at finite $T$
with a slope proportional to $T$.

\begin{figure}[t]
\includegraphics[width=1.0\linewidth,height=0.72\linewidth]{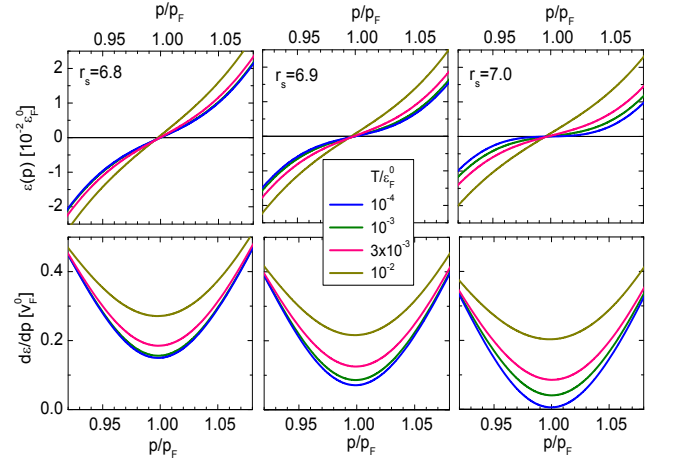}
\hskip 1.5 cm
\caption{Single-particle spectrum $\epsilon(p)$
(top panels) and derivative $d\epsilon(p)/dp$ in units of
$v_F^0=p_F/M$ (bottom panels), evaluated in the model
(\ref{model_2pf}) with parameters $g_2=-0.16$ and $\beta_2=0.14$,
chosen to adequately describe results of microscopic calculations
\cite{bz} of 2D-electron-gas single-particle spectrum at $T=0$,
are shown as functions of $p/p_F$ at four (line-coded)
temperatures expressed in units of $10^{-2}\,\varepsilon_F^0$ at
$r_s=6.8$ (left column), $r_s=6.9$ (middle column), and $r_s=7.0$
(right column).}
\label{fig:eg_fl}
\end{figure}

On the other hand, in the state having a multi-connected Fermi
surface, the spectrum $\epsilon(p,T=0)$ varies smoothly
in the space beyond the sheets of the Fermi surface but oscillates
rapidly in the space between them.  The magnitude of the
departure of $|\epsilon(p,T=0)|$ from 0 in this domain, denoted below by
$T_m$, emerges as a new energy scale of the problem, at which ``melting''
of the structure associated with the well-defined multi-connected
Fermi surface can occur. Indeed, as seen from the basic Landau
formula (\ref{dist}), the distribution $n(p,T=0)$ remains almost
unchanged as long as $T<T_m$.  However, as the temperature $T$ increases,
kinks in $n(p)$ become smeared, and at $T>T_m$, in the momentum domain
between the sheets, the function $n(p,T)$ becomes continuous and almost
$T$-independent. In this case, according to Eq.~(\ref{spte}), the
dispersion of the single-particle spectrum $\epsilon(p)$ becomes
proportional to $T$.  {\it Thus at $T>T_m$, the temperature
evolution of $\epsilon(p,T)$ and $n(p,T)$ becomes universal.}

\begin{figure}[t]
\includegraphics[width=1.0\linewidth,height=1.\linewidth]{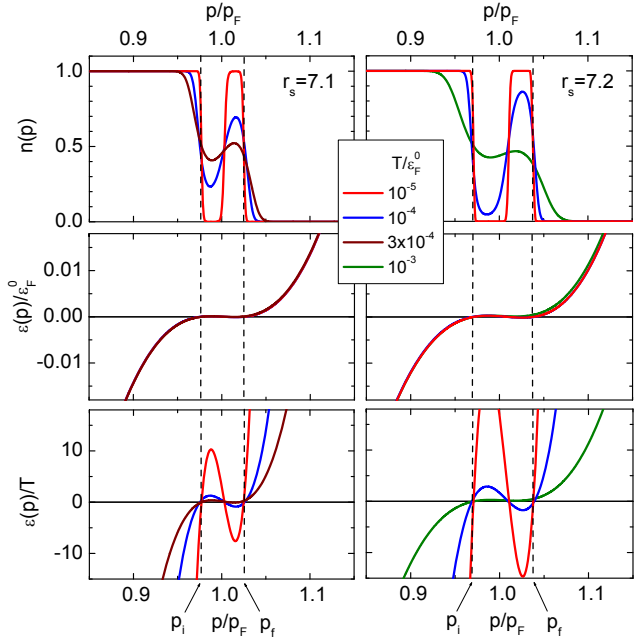}
\hskip 1.5 cm
\caption{Occupation numbers $n(p)$ (top panels),
single-particle spectrum $\epsilon(p)$ in units of $\epsilon_F^0$
(middle panels), and ratio $\epsilon(p)/T$ (bottom panels) for the
model (\ref{model_2pf}) at $r_s=7.1$ (left column) and $r_s=7.2$
(right column), exceeding the QCP value $r_{\infty}=7.0$, evaluated with the same set of the parameters as in Fig.~\ref{fig:eg_fl}. All three quantities are shown as functions of $p/p_F$ at different temperatures in units of $\varepsilon_F^0$, lower than the transition temperature $T_m=10^{-3}\varepsilon^0_F$.
}
\label{fig:eg_b}
\end{figure}

\begin{figure}[t]
\includegraphics[width=0.95\linewidth,height=1.\linewidth]{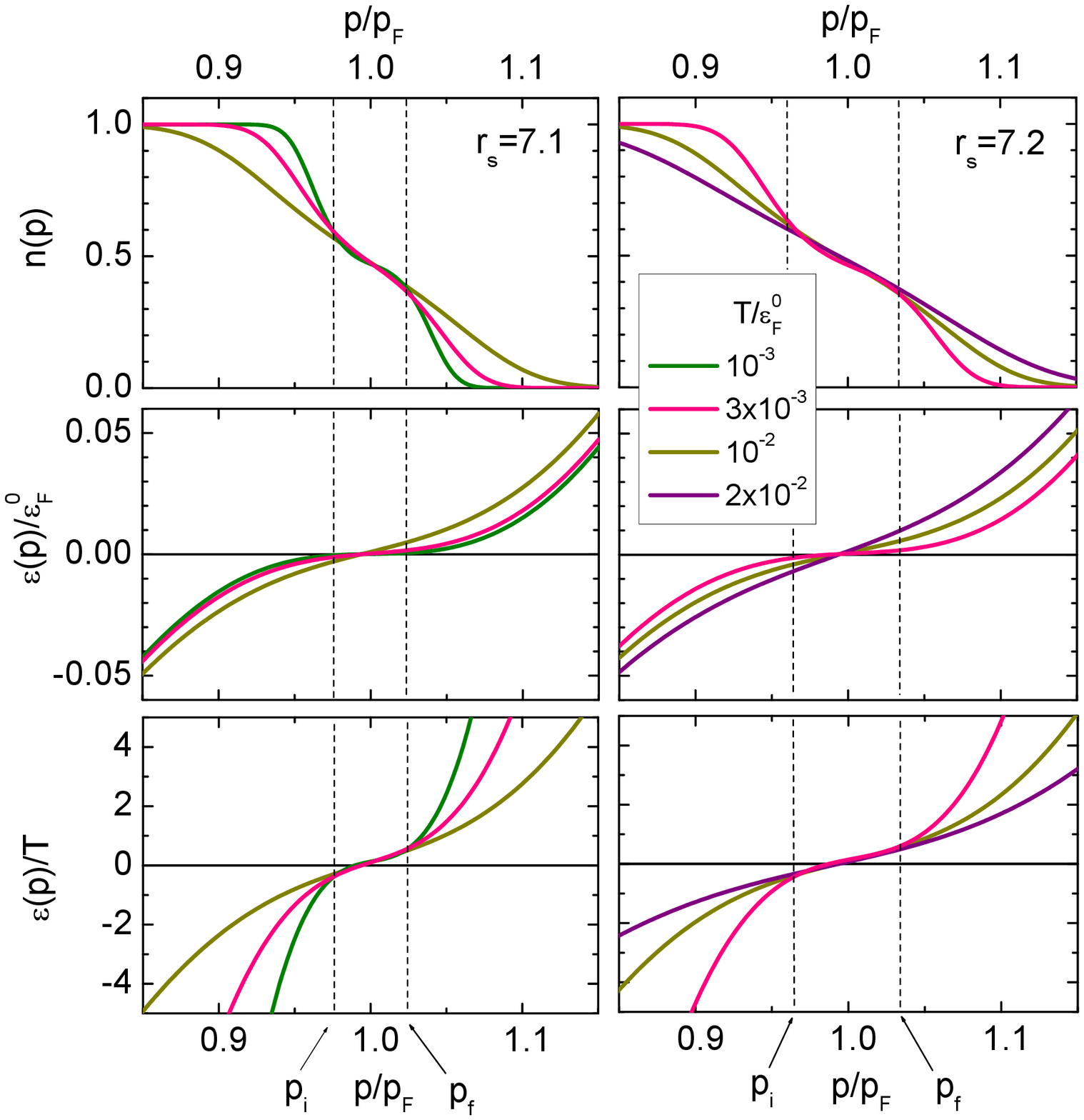}
\hskip 1.5 cm
\caption{Same as in Fig.~\ref{fig:eg_b} but at temperatures, higher than $T_m$.}
\label{fig:eg_fc}
\end{figure}

\begin{figure}[t]
\includegraphics[width=0.65\linewidth,height=1.2\linewidth]{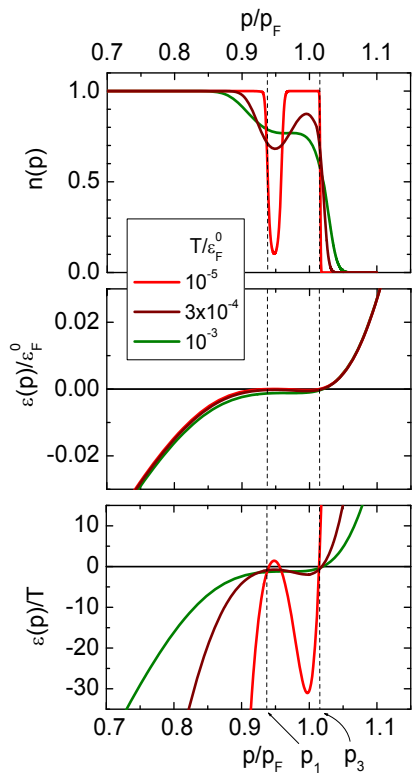}
\hskip 1.5 cm
\caption{Occupation numbers $n(p)$ (top panel), single-particle spectrum $\epsilon(p)$ in units of $10^{-3}\,\varepsilon_F^0$ (middle panel), and ratio $\epsilon(p)/T$ (bottom panel), plotted versus $p/p_F$ at three line-type-coded temperatures in units of $\varepsilon_F^0$, taken below the transition temperature $T_m=3\times 10^{-3}\,\varepsilon^0_F$. The model (\ref{model_0pf}) is assumed with parameters $g_3=0.45$ and $\beta_3=0.07$.}
\label{fig:b_0pf}
\end{figure}

\begin{figure}[t]
\includegraphics[width=0.65\linewidth,height=1.2\linewidth]{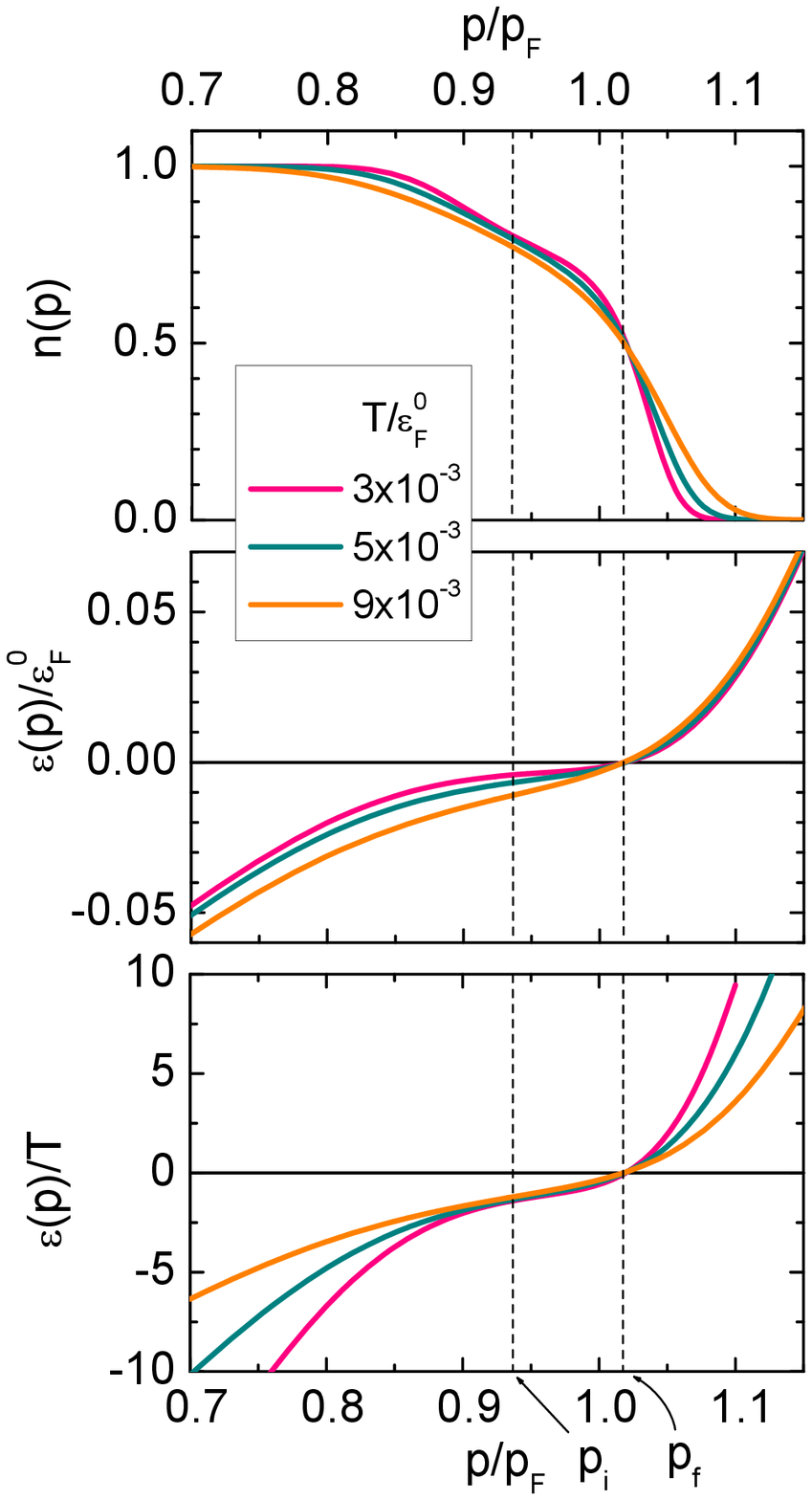}
\hskip 1.5 cm
\caption{Same as in Fig.~\ref{fig:b_0pf} but at temperature higher than $T_m$.}
\label{fig:fc_0pf}
\end{figure}

\begin{figure}[t]
\includegraphics[width=0.65\linewidth,height=1.2\linewidth]{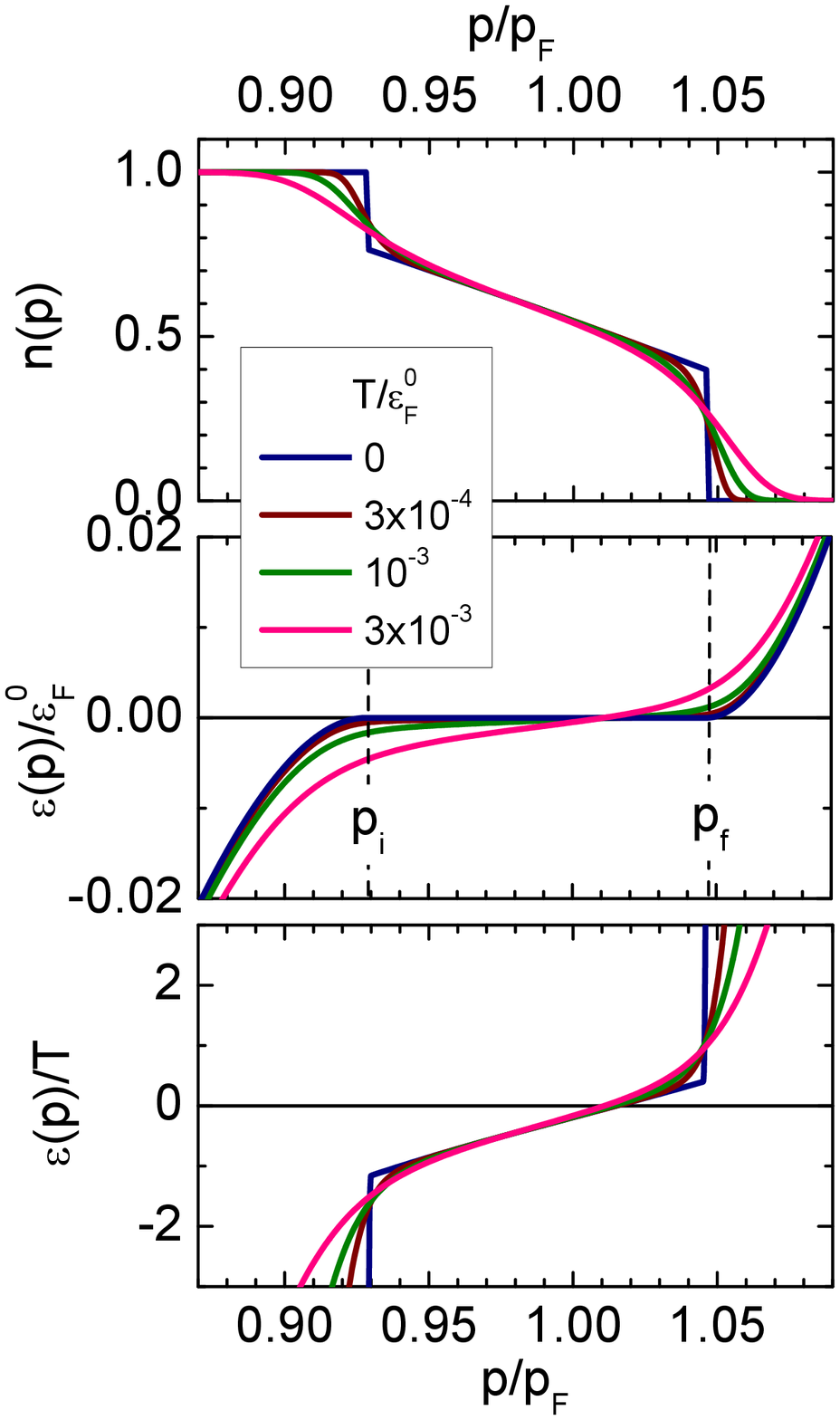}
\hskip 1.5 cm
\caption{Single-particle spectrum $\epsilon(p)$ in units of $\varepsilon_F^0$ (top panels), occupation number $n(p)$ (middle panels), and ratio $\epsilon(p)/T$ (bottom panels) for the model (\ref{model_yuka}) with the parameters $\beta_*=10$ and $g_*=70.0$. All three quantities are shown as functions of $p/p_F$ at different line-type-coded temperatures in units of $\varepsilon_F^0$.}
\label{fig:fc_yuka}
\end{figure}

To provide tangible support for this important conclusion,
Figs.~\ref{fig:eg_fl}--\ref{fig:fc_0pf} present the results
of model calculations of the spectrum $\epsilon(p,T)$ and momentum
distributions $n(p)$ performed in Ref.~\onlinecite{prb2008} on
the basis of Eq.~(\ref{lansp}).  Two different interaction
functions $f$ are employed in these calculations, namely
\beq
  f(k)=g_2{\pi\over M} {1\over (k^2/4p_F^2-1)^2+\beta^2_2},
\label{model_2pf}
\eeq
with $g_2=-0.16$ and $\beta_2=0.14$ and
\beq
  f(k)=g_3{\pi^2 p_F\over M}{1\over k^2+\beta^2_3p^2_F}.
\label{model_0pf}
\eeq
The first is chosen because this function adequately reproduces,
within the extended Landau-Migdal treatment, the results of microscopic
calculations \cite{bz} of the single-particle spectrum $\epsilon(p)$
of the 2D electron gas at $T=0$, and because the corresponding
initial TPT essentially coincides with the QCP located at
$r_s=r_{\infty}=7.0$.  The second choice of $f$ is relevant
to a 3D system with spin fluctuations and has already been
considered above in Sec.~VII.B.

Fig.~\ref{fig:eg_fl} displays results for the single-particle
spectrum $\epsilon(p)$ and the group velocity $d\epsilon(p)/dp$,
obtained on the FL side of the QCP for the model (\ref{model_2pf}).
It is seen that as the QCP is approached, the group velocity becomes
a parabolic function of momentum $p$.  With increasing $r_s$ its
minimum gradually approaches the horizontal axis, touching down for
$r_s=r_{\infty}$ exactly at the point $p=p_F$, so that
$\epsilon(p,T=0)\sim (p-p_F)^3$ has an inflection point at
the Fermi surface.  At $r_s>r_{\infty}$, the group velocity $v_F$
evaluated with $n(p)=n_F(p)$ changes its sign, triggering
the rearrangement of the Landau state, in accordance
with the topological scenario for the QCP.

We turn now to Figs.~\ref{fig:eg_b} and \ref{fig:eg_fc}, which show
numerical results for the spectrum $\epsilon(p)$ and quasiparticle
momentum distribution $n(p)$, calculated for the interaction
function (\ref{model_2pf}) both below and above the temperature
$T_m$. It should be noted that the value of $T_m$ is somewhat
uncertain, since we are dealing here with a crossover rather than
a well-defined phase transition.
Nevertheless, comparison of these two figures reveals a striking
alteration of the structure of both $\epsilon(p)$ and $n(p)$ upon
passage across $T_m$ from the lower to the higher temperature.
Indeed, as seen from the top panel of Fig.~\ref{fig:eg_b}, a
well-defined multi-connected Fermi surface, distinguished by a
pronounced gap in filling, exists only at extremely low
temperatures $T< 10^{-4}\varepsilon^0_F$, while at $T\simeq
T_m\simeq 10^{-3}\varepsilon^0_F$, the gap in the occupation
numbers closes, although at $T<T_m$, some $T$-dependence of $n(p)$
in the vicinity of the Fermi surface is still present.  On the
other hand, as seen from the top panel of Fig.~\ref{fig:eg_fc}, a
$T$-independent behavior of the momentum distribution $n(p)$,
existing at $T>T_m$ in the FC domain, holds over a wide temperature
interval.  Comparison of the third panels of Figs.~\ref{fig:eg_b}
and \ref{fig:eg_fc} demonstrates that huge variations of the ratio
$\epsilon(p)/T$, existing at $T<T_m$, completely disappear in the
FC domain at $T>T_m$.

The same conclusions follow from Figs.~\ref{fig:b_0pf} and
\ref{fig:fc_0pf}, which display the temperature behavior of the
spectrum $\epsilon(p)$ and momentum distribution $n(p)$, evaluated
for the interaction function (\ref{model_0pf}) below and above
$T_m$.

It is instructive to compare the results shown in
Figs.~\ref{fig:eg_b}--\ref{fig:fc_0pf}
with corresponding results derived for the singular interaction
function
\beq
  f(k)=g_*{\pi^2\over M}{e^{-\beta_* k/p_F}\over k},
\label{model_yuka}
\eeq
studied in Ref.~\onlinecite{physrep} as a model of FC analytically
soluble at $T=0$.  (All the input parameters in the interactions
defined in Eqs.~(\ref{model_2pf}), (\ref{model_0pf}), and
(\ref{model_yuka}) are dimensionless.)

It is characteristic of systems with a singular interaction function
(singular as $k\to 0$, hence of long range in coordinate space)
that the linear-in-$T$ dispersion of the single-particle spectrum
symptomatic of the FC domain exists at any temperature $T$.  This
behavior is evident in the numerical results for the interaction model
(\ref{model_yuka}) plotted Fig.~\ref{fig:fc_yuka}.

From these and other numerical and analytical studies, it may
be concluded that specific features of the flattening of single-particle
spectra beyond the first TPT point are {\it universal}, differing only
with respect to the scale temperature $T_m$.

\section{Unconventional thermodynamics of systems with a FC}

In this section we will focus on certain unusual features of
the thermodynamic properties of strongly correlated systems hosting
a FC. In making concrete estimations, it will be assumed that the
interaction function has a small, specific fluctuation contribution,
the origin and form of which has been analyzed in Sec.VII.B.

\subsection{Logarithmic corrections to $\gamma(T)$}

The standard FL behavior of $\gamma(T\to 0)={\rm const.}$ does
in fact re-emerge beyond the point of a TPT, because the density
of states $N(0)$ again becomes finite. However, recovery of FL
behavior occurs only at extremely low temperatures. We will see
that, as the temperature increases, $\gamma$ acquires a logarithmic
correction $\delta \gamma_{\fl}(T)\propto \ln (1/T)$.  Such a
correction is in fact observed in a number of heavy-fermion metals
at temperatures exceeding a rather low temperature $T_0$.  Its
presence is usually explained within the framework of the collective
scenario for the QCP.  As already discussed, in this scenario
quasiparticle weight $z$ vanishes at a second-order transition
point.  Consequently, the mass operator $\Sigma(p,\varepsilon)$
contains a marginal term $\sim \varepsilon\ln\varepsilon$,
giving rise to the appearance of a logarithmic correction
$\delta \gamma_{\fl}(T)\propto \ln (1/T)$ to the FL Sommerfeld
ratio $\gamma_{\fl}(T)={\rm const}$.

However, as we will be readily appreciated, breakdown of the
quasiparticle picture  is not a necessary condition for the emergence of
logarithmic corrections to the Sommerfeld ratio $\gamma(T)$.
Such corrections can emerge in the QCP region due to the presence
of small spin-fluctuation contributions to the FL interaction
function.  To see this, let us evaluate $\gamma(T)$ within
the spin-fluctuation scenario introduced in Sec.~VII.B. One has
\beq
\gamma(T)\sim T^{-1}
\int \epsilon(p){\partial n(p,T)\over \partial T}d\upsilon,
\label{som}
\eeq
where $n(p,T)=\left(1+\exp(\epsilon(p,T)/T)\right)^{-1}$
is the usual Landau quasiparticle occupation number.  Referring
to Eq.~(\ref{vsf}) we know already that at $\xi^{-1}\ll (p-p_F)\ll p_F$,
the group velocity $v(p)$ takes on a logarithmic correction
$\sim \ln\left((p-p_F)/p_F\right)\sim \ln \left(M^*\epsilon/p_F\right)$,
implying that
\beq
{1\over v(\epsilon)}\simeq {1\over v_F} \left(1-{3g^2\over 2\pi v_F}
\lambda\ln \left({M^*\epsilon\over p_F}\right)\right).
\label{1v}
\eeq
Since the overwhelming contributions to $\gamma(T)$ and other
thermodynamic quantities come from the energy region where
$|\epsilon|\simeq T$, we introduce a new variable $y=\epsilon/T$
to obtain the standard FL formula
\beq
\gamma(T,\rho)\sim\int y^2 n(y)(1-n(y)){1\over v(y)}d y.
\label{ct}
\eeq
Upon inserting Eq.~(\ref{1v}) into
this equation, we infer that the logarithmic correction to
$\gamma_{\fl}={\rm const.}$ becomes well-pronounced already at a rather
low temperature $T_0=T_F/(p_F\xi)$, with $T_F=p^2_F/2M^*$, according
to
\beq
{\delta \gamma_{\fl}(T)\over \gamma_{\fl}}\simeq {3g^2M^*
\over \pi p_F}\ln\left({T_F\over T}\right), \quad
  T_0<T<T_F    .
\label{som1}
\eeq
The $T$-dependence of this correction coincides with that obtained within
the conventional scenario for the QCP.  But importantly: in contrast
to this scenario, the regime involved is {\it far from} the critical one
where the quasiparticle picture breaks down.  Similar logarithmic
corrections emerge in the same temperature interval when evaluating
the thermal expansion coefficient, magnetic susceptibility, and other
thermodynamic quantities.  Their presence is irrelevant to
violation of the quasiparticle picture.

The value of the parameter $(p_F\xi)$, and hence that of the governing
parameter $\lambda$, can be extracted from experimental data
obtained on the upper and lower boundaries of the interval where
the logarithmic behavior (\ref{som1}) is in effect. The analysis
of these data yields $10\leq p_F\xi\leq 10^2 $, implying that
\beq
10^2<\lambda<10^4.
\label{ineq}
\eeq
Comparing the Sommerfeld ratios on opposite edges of the interval
of logarithmic behavior, one can also estimate the value of another
parameter, namely $g^2/(\pi v^0_F)$.  Since $\gamma$ drops by
factor 2 or so, we conclude that $g^2/(\pi v^0_F)<1/\ln(p_F\xi)<0.1$.

These formulas, in first turn, Eq.~(\ref{ct}), are valid provided
the inequality $T_0<T_m$ is met.  Otherwise, at $T>T_m$ the FC forms,
leading to a dramatic change of the Sommerfeld ratio $\gamma(T)$.
Indeed, according to the foregoing analysis, the major part of
the FC momentum distribution is $T$-independent and therefore does
not contribute to $\gamma(T)$.  Consequently, when formation of
the FC is completed, the enhancement factor in $\gamma(T)$
disappears in spite the huge FC density of states.

  \subsection{Entropy excess}

A drastic change in the behavior of the entropy $S(T)$ occurs
\cite{ks,physrep,yak} when, at $T>T_m$, a FC is formed in the
domain $\cal C$ under rearrangement of the momentum distribution
$n(p)$ into the self-consistent FC solution $n_*(p)$.  The basic
entropy formula (\ref{entr}) of the original quasiparticle formalism
remains intact:
\beq
S_*=-2\int\!\!\int [n_*({\bf p})\ln n_*({\bf p})
+(1-n_*({\bf p}))\ln (1-n_*({\bf p}))] d\upsilon.
\label{entrfc}
\eeq
However, due to the NFL component in $n_*(p)$, the system is seen
to possess a {\it $T$-independent} entropy excess $S_*(\rho)$.  The
situation we now face---with the strongly correlated fermion system
having a finite value $S_*$ of the entropy at $T\to 0$---resembles
that encountered in a system of localized spins. In the spin system,
the entropy referred to one spin is simply $\ln2$, while in the
system having a FC, we have $S_*/N\simeq\eta \ln2$, where
$\eta=(p_f-p_i)/p_F$ is the dimensionless FC parameter.

On the other hand, numerical calculations demonstrate that within
the FC domain, the momentum distribution $n_*(p)$ changes rapidly
under variation of the total density $\rho$.  The corresponding
nonzero value of the derivative $\partial S_*/\partial\rho$ produces
a huge enhancement of the thermal expansion coefficient
$\beta\sim\partial S(T)/\partial\rho\simeq \eta$ with respect to
its FL value, proportional to $T$.\cite{zksb} Consonant with
this result, in many heavy-fermion metals it is found that
$\beta$ is indeed almost temperature-independent and that it
exceeds typical values for ordinary metals by a factor
$10^3$--$10^4$.\cite{oeschler}

\subsection{Curie-Weiss behavior of the spin susceptibility}

Another peculiar feature of strongly correlated Fermi systems in
the QCP region involves the temperature dependence of the spin
susceptibility $\chi(T)=\chi_0(T)/(1+g_0\Pi_0(T))$, where
$$
\chi_0=\mu^2_e\Pi_0(T)=-2\mu^2_e\int {dn(p,T)
\over d\epsilon(p)}d\upsilon
$$
\beq
=2{\mu^2_e\over T}
\int n(p,T)(1-n(p,T))d\upsilon
\label{chio}
\eeq
and $g_0$ is the spin-spin component of the interaction function.
As mentioned before, in the QCP region, the standard FL structure is
preserved only at $T<T_m$. In particular, the standard Pauli behavior
$\chi(T)={\rm const.}$ is maintained until ``melting'' of
the multi-sheet structure occurs at $T\simeq T_m$, giving
rise to flattening of the single-particle spectrum associated
with a FC.

At $T>T_m$, insertion of $n_*(p)$ into Eq.~(\ref{chio}) yields
the Curie-like term
\beq
\chi_*(T)=\mu^2_e{ C_{\mbox{\scriptsize eff}}(\rho)\over T}
\eeq
with an effective Curie constant
\beq
C_{\mbox{\scriptsize eff}}(\rho)=2\int n_*(p)(1-n_*(p))d\upsilon
\label{ceff}
\eeq
that depends dramatically on the density.\cite{zk4,yak}
Since $C_{\mbox{\scriptsize eff}}$ is proportional to
the FC parameter $\eta$, we infer that
\beq
C_{\mbox{\scriptsize eff}}\simeq S_*.
\eeq
Thus, all compounds in which the spin susceptibility exhibits the
Curie-like behavior possess a large entropy.  Furthermore, in the
whole temperature interval from $T=0$ to $T>T_m$, the spin
susceptibility of a Fermi system beyond the QCP possesses
Curie-Weiss-like behavior $\chi(T)\sim 1/(T-T_W)$ with a {\it negative}
Weiss temperature $T_W$.  Measurements in $^3$He films on various
substrates and in numerous heavy-fermion compounds provide examples of
this NFL behavior. We emphasize that in our scenario, the negative
sign of $T_W$ holds even if the spin-spin interaction is attractive.
This contrasts with the behavior in a system of localized spins,
where the Weiss temperature has a negative sign only if the
spin-spin interaction is {\it repulsive}.  Moreover, in the
case of localized spins, the Stoner factor must be suppressed,
whereas in the vicinity of the QCP, this factor is usually enhanced.

Another conspicuous feature of the physics beyond the
QCP is associated with the Sommerfeld-Wilson ratio
$R_{\rm SW}=T\chi(T)/\mu^2_eC(T)\sim S_*/C(T)$.  Since the excess
entropy $S_*$ does not depend on $T$, it makes no contribution
to the specific heat $C(T)$; consequently in systems with a FC,
one should see a great enhancement of $R_{\rm SW}$.

\section{Classical behavior near the QCP}

In previous sections, we have discussed the structure of the extended
quasiparticle picture near the QCP and beyond it. In this section,
we highlight one of the most distinctive (and counterintuitive)
features of strongly correlated Fermi systems in this domain of the
Lifshitz phase diagram, notably their classical behavior.  This
aspect was first revealed in measurements of the specific
heat $C(T)$ of dense $^3$He films at the lowest temperatures
$T\simeq 1$ mK reached experimentally.  As the temperature is
lowered into this regime, the specific specific heat $C(T)$
behaves as \cite{saunders1,saunders2}
\beq
C(T)=\beta+\gamma T,
\label{c2d}
\eeq
thus exhibiting an unexpected $T$-independent term $\beta(\rho)$.
Such behavior contrasts sharply with that of its 3D counterpart,
bulk liquid $^3$He, for which $C(T)$ obeys standard FL theory
in approaching zero linearly in $T$, with a prefactor $\gamma$
proportional to the density of states $N(0)\propto M^*$.  Indeed,
FL theory works better and better as $T$ is lowered toward
zero, provided no superfluid transition intervenes.

It is has been commonly assumed \cite{golov} that the term $\beta(\rho)$
is related to the nature of the substrate which supports the 2D $^3$He
film.  More specifically, it is thought that due to weak heterogeneity
of the substrate (steps and edges on its surface), quasiparticles
are delocalized from it, giving rise to the $\beta$ feature of the
heat capacity.\cite{saunders2}  This explanation is undermined
somewhat by the fact that if substrate-induced disorder
were responsible for the $\beta$ feature, one would expect the
departure of $C(T)$ from the FL prediction to decrease as the film
density increases, since the effects of disorder are most pronounced
in weakly-interacting systems.  Contrariwise, the specific-heat anomaly
makes its appearance in just that density region where the effective mass
$M^*$ is enhanced and 2D liquid $^3$He system becomes strongly
correlated.\cite{saunders1,saunders2}  In this situation,
the impact of disorder should be suppressed.  The weight of these
considerations compells us to treat the observed behavior
of $C(T)$ as an {\it intrinsic} property of 2D liquid $^3$He.

Curiously, due to the presence of the residual term $\beta$, this
behavior mimics the classical Dulong-Petit (DP) Law of the
specific heats $C(T)$ of solids, the major contribution to which
comes from phonons, being obtained from the formula
\beq
C_B(T)\propto\int\limits_0^{k_{\rm max}}
\omega(k) {\partial n_B(k)\over \partial T} {kdk\over 2\pi},
\label{textf}
\eeq
where $n_B(k)=[e^{{\omega(k)/T}}-1]^{-1}$ is the Bose-Einstein
phonon distribution function.  The $T$-dependence of $C_B(T)$
is determined by relation between the temperature $T$ and the
Debye temperature $\Omega_D= \omega(k_{\rm max})\simeq ck_{\rm max}$.
At $T\geq \Omega_D$, the specific heat becomes independent of $T$,
and the DP behavior is recovered.  The greater the value of the parameter
$k_{\rm max}$ characterizing the cutoff of the phonon spectrum,
the more pronounced is the classical contribution $\beta$ to
the specific heat $C(T)$, which is proportional to $k^D_{\rm max}$
for a system of dimensionality $D$.

Normally, $\Omega_D$ is sufficiently high that, true to its
empirical discovery, the DP Law belongs to the realm of classical
physics.  However, we will see that in strongly correlated Fermi
systems, the characteristic frequency of some collective mode
can be extremely small such as to allow DP behavior (\ref{c2d})
in 2D liquid $^3$He at millikelvin temperatures.  Certainly,
neither the usual hydrodynamic sound nor longitudinal zero sound
can qualify, since the group velocities of both modes remain
finite as $M^*\to \infty$.\cite{pines,halat}  The spin fluctuation
mode is excluded as well, however, due to substantial Landau damping.
The desired mode is in fact provided by transverse zero sound
mode (TZSM),\cite{dplett} whose properties near the QCP can
be explicated in terms of its well-known dispersion relation
in 3D matter \cite{halat}
\beq
{s\over 2}\ln {s+1\over s-1}= {F_1-6\over 3F_1(s^2-1)},
\label{dt}
\eeq
conveniently rewritten as
\beq
(s^2-1)\left({s\over 2}\ln {s+1\over s-1}-1\right)
={1\over 3}-{2\over F_1},
\eeq
with $s=c/v_F$ and $F_1=f_1p_FM^*/\pi^2$.  The TZSM is seen
to propagate only if $F_1>6$, i.e., provided $M^*>3M$.  Near the
QCP where $M^*(\rho)\to\infty$, this requirement is always
met.  In this case, one has $v_F/c\to 0$, and Eq.~(\ref{dt})
simplifies to
\beq
1={F_1\over 15}{v^2_F\over c^2},
\label{dr11}
\eeq
which implies
\beq
c(\rho\to \rho_{\infty})\to \sqrt{{p_Fv_F(\rho)\over  M}}
\propto {p_F\over M}\sqrt{{M\over M^*(\rho)}}\to 0,
\label{ctr}
\eeq
an analogous formula being obtained for a 2D system.

It should now be clear that toward the QCP, the effective
Debye temperature $\Omega_t=\omega(k_{\rm max})=
c_Rk_{\rm max}$ {\it goes down to zero}, independently of the
value of the wave number $k_{\rm max}$ corresponding to the
saturation of the TZSM spectrum.  Thus, the necessary condition
$\Omega_t<T$ for emergence of a regime of classical behavior
is always met.  However, another condition must also be satisfied
if there is to exist a well-pronounced classical domain at
extremely low temperature: the parameter $k_{\rm max}$ must not
to be too small.

There is no such a restriction in the conventional electron-phonon
problem. Indeed, the phonon group velocity $c$ depends weakly on
the wave number $k$, whose characteristic value $k_{\rm max}$
coincides with its maximum  possible value $\simeq 1/r_0$, the
inverse distance between particles in the system. As a result, the
phonon contribution to the specific heat $C(T)$ turns out to be
proportional to the particle density $\rho$.  However, softening
of the TZSM is terminated around some rather small critical wave
number $k_{\rm max}$, so that at greater wave numbers, $\omega(k)$
becomes larger than $T$, destroying classical behavior.  Evaluation
of the corresponding threshold value is a difficult problem,
because its solution requires analysis beyond the long-wave
approximation.  Here we consider the situation in systems with
only a small proportion of FC, where this problem can be handled
with some facility based on the familiar FL kinetic
equation \cite{trio,pines}
\beq
\left(\omega-{\bf k}{\bf v}\right)\delta n({\bf  p})=
-{\bf k}{\bf n}{\partial n(p)\over \partial p}
\int {\cal F}({\bf p},{\bf p}_1) \delta n({\bf p}_1)d\upsilon_1.
\label{kin}
\eeq
Focusing on transverse zero sound in 2D liquid $^3$He, we need
only to include the term in the Landau amplitude ${\cal F}$ involving
the first harmonic $f_1$.  Making the usual identification
$(c_t-\cos\theta)\delta n({\bf p})=\left(\partial n(p)/\partial
p\right)\phi({\bf n})$, where $\cos\theta={\bf k}{\bf v}/kv$,
Eq.~(\ref{kin}) becomes
\beq \phi(\theta) =
-f_1p_F\cos\theta\int \cos\chi {\partial n(p_1)/\partial
p_1\over c_t-v(T)\cos\theta_1}\phi(\theta_1) {dp_1d\theta_1\over
(2\pi)^2}\label{kins},
\eeq
where $\cos\chi= \cos\theta\cos\theta_1+\sin\theta\sin\theta_1$, while
$v(T) $ is the FC group velocity, proportional to $T$.  The solution
describing transverse zero sound is $\phi({\bf n})\sim \sin\theta\cos\theta$.

It is seen immediately that $c_t\gg v(T)\sim T$; therefore the transverse
sound in question does not suffer Landau damping.  Then, upon retaining only
the leading relevant term $v(T)\cos\theta/c^2_t$ of the expansion
of $1/(c_t-v(T)\cos\theta)$ and performing straightforward
manipulations, we are led to the simple result
\beq
c^2_t\propto -{p_F\over M} \int {\partial n(p)\over \partial p} v(p,T)dp.
\eeq
Factoring out an average value of the group velocity $v(p,T)\simeq T/p_F$,
we find that
\beq
c_t
\propto \sqrt{{T\over M}}
\label{ctrs},
\eeq
i.e., in the FC domain of the phase diagram, the velocity
$c_t$ of the transverse mode {\it depends on temperature}
so as to vanish like $\sqrt{T}$ as $T \to 0$.

We now demonstrate that such a softening of the TZSM holds only
as long as the wave number $k$ does not exceed
$k_{\rm max}\simeq L=p_f-p_i$.  The reason for this
termination is that the {\it noncondensed} component of the
quasiparticle system in 2D liquid $^3$He comes into play at
$k>L$, consisting of quasiparticles with normal $T$-independent
dispersion $d\epsilon(p)/dp\simeq p_F/M$.  The group velocity
$c_t(k>L)$ of the transverse mode then soars upward, rendering
the corresponding contribution to $C(T)$ irrelevant.

To evaluate this effect it is necessary to go beyond the
long-wave approximation, a process beset with replacement of
${\bf k}{\bf n} \partial n(p)/\partial p$ by
$n({\bf p}+{\bf k})-n({\bf p})$ and ${\bf k}{\bf v}$ by
$\epsilon({\bf p}+{\bf k})-\epsilon({\bf p})$.  With these
replacements, simple algebra leads to the behavior
\beq
\omega^2(k)\propto-\int \left(n({\bf p})-n({\bf p}+{\bf k})\right)
\left(\epsilon({\bf p})-\epsilon({\bf p}+{\bf k})\right) d\upsilon.
\label{tzfl}
\eeq
At $k\leq L$, almost all the FC states contribute to this
expression on an equal footing, yielding relation (\ref{ctrs}). However,
at $k > L$, the predominant contributions to the integral {\ref{tzfl}}
come from momentum regions where the difference
$|\epsilon({\bf p})-\epsilon({\bf p}+{\bf k})|$ has its maximum value,
which is $T$-independent.  Thus we infer that at such large $k$ values,
softening of the spectrum $\omega_t(k)$ is terminated---in contrast
to our previous claim \cite{dplett} that softening of $\omega_t(k)$
persists until $k>\sqrt{p_FL}$.

Since, as we have seen, $k_{\rm max}\simeq L$.  Hence we arrive at
the result
\beq
\Omega_t\simeq k_{\rm max}c_t\propto L\sqrt{{T\over M}}.
\eeq
As long as the inequality $L<(MT)^{1/2}$ holds (or equivalently,
$T/\epsilon^0_F >L/p_F$), the ratio $\Omega_t/T$ remains small, and
the Dulong-Petit law $C(T)={\rm const.}$ obtains.  Thus, in spite of
the low temperature, the specific heat behaves as if the system were
situated in the classical regime.  This paradoxical outcome is a
consequence of the presence, in the strongly correlated Fermi system,
of a macroscopic subsystem with heavy quasiparticles.  As the
temperature ultimately goes down to zero at a fixed density $\rho$, the
inequality $L < (MT)^{1/2}$ eventually fails, the quantum regime
is restored, and the dominant contribution to the specific heat
comes from the ``normal'' fermions.  In other words, by reducing
$T$ sufficiently, one can reach a domain in which the FL behavior
of the specific heat $C(T)$ is recovered.

Interestingly, the value of the constant term in $C(T)$ can be
evaluated in closed form in terms of the FC range $L$.  Upon
inserting $\omega_t(k)=c_tk$ into Eq.~(\ref{textf}) and
integrating, the $T$-independent term in the specific heat
is found to be
\beq
{C\over N}= {L^2\over 8\pi\rho},
\label{crel}
\eeq
where $N$ is the number of atoms in the film.

We turn finally to a discussion of the impact of the TZSM on
transport properties.  In the case of a convention Fermi liquid,
the Fermi surface consists of a single sheet, so the TZSM has
a single branch with velocity $c_t$ exceeding the Fermi velocity
$v_F$.  Consequently, emission and absorption of sound quanta by
electrons is prohibited, and the role of the TZSM in kinetics is
of little interest.  However, in heavy-fermion metals, it is
common for several bands to cross the Fermi surface simultaneously,
thereby generating several zero-sound branches.  For all branches
but one the sound velocities are less than the largest Fermi
velocity.   Hence the aforementioned ban is lifted, and these
branches of the TZSM spectrum do experience damping, in a situation
similar to that for zero-spin sound.  In the latter instance,
Landau damping is so strong that the mode cannot propagate through
the liquid.\cite{pines,halat}  It will be seen that this is {\it not}
the case for damping of the TZSM, because of the softening of
this mode close to the QCP.  Due to the softening effect, the
contribution of the damped TZSM to the collision integral has
the same form as the electron-phonon interaction at room temperature.

To facilitate analysis of damping of the TZSM in systems having
a {\it multi-connected} Fermi surface, we restrict consideration
to the case of two electron bands.  The TPT is assumed to occur at
one of the bands, so that its Fermi velocity, denoted again by
$v_F$, tends to zero, while the Fermi velocity $v_o$ of the other
band remains unchanged through the critical density region.  The
model dispersion relation for the complex sound velocity $c = c_R+ic_I$
becomes
$$
1={F_1\over 6}\left[ 1-3\left({c^2\over v^2_F}-1\right)
\left({c\over 2v_F}\ln {c+v_F\over c-v_F}-1\right)\right]+$$
\beq
+{F_1\over 6}{v_F\over v_o}\left[1-3\left({c^2\over v^2_o}-1\right)
\left({c\over 2v_o}\ln {c+v_o\over c-v_o}-1\right)\right].
\label{dr2}
\eeq
It can easily be verified that the contribution of the second term to
the real part of the right-hand side of Eq.~(\ref{dr2}) is small
compared to that of the first term, since $v_F/v_o\to 0$ toward the QCP.
On the other hand, noting that
$\ln \left[(c_R+ic_I+v_o)/ (c_R+ic_I-v_o)\right] \simeq - i \pi$,
the corresponding contribution  $i\pi F_1v_Fc_R/(4v^2_o)$
to the imaginary part of the right-hand side cannot be ignored,
else $c_I=0$.  By this reasoning, Eq.~(\ref{dr2}) assumes the
simplified form
\beq
1={F_1\over 15}{v^2_F\over (c_R+ic_I)^2}- i{\pi\over 4v^2_o} F_1 v_Fc_R
\eeq
analogous to Eq.~(\ref{dr11}).  Its solution obeys
\beq
c_R\propto \sqrt{{M\over M^*(\rho)}}, \quad c_I\propto {M\over M^*(\rho)}.
\label{rvel}
\eeq
Importantly, we see then that the ratio $c_I/c_R\propto\sqrt{M/M^*(\rho)}$
is {\it suppressed} in the QCP regime, which allows us to analyze
the contribution of the TZSM to the collision term entering the resistivity
along the same lines as in the familiar case of the electron-phonon
interaction.

\section{Conclusion}

Proceeding from the original FL quasiparticle picture due to
Landau and Migdal, we have addressed
the formation of flat bands in
strongly correlated Fermi systems beyond a point where the necessary
condition for stability of the Landau state is violated and hence
subject to rearrangement.
Responding to a soaring exhortation from Migdal, ``{\it Beri shire!}''
(Embrace everything you can!), expressed to an interlocutor when
A.~B.\  approved what he heard, we have analyzed
 this phenomenon in
diverse strongly correlated Fermi systems from neutron stars to atomic liquids
 to electron systems of solids.  Absent, however, are reviews of
the latest achievements in the investigation of flattening of
single-particle spectra in equations of particle physics,\cite{lee}
in topological media,\cite{vol2010a,vol2010b,vol2010c} including
the analysis of flat bands on the surface of  multi-layered graphene (see Refs.~\onlinecite{neto1,neto2} and references therein).

We have seen that in the class of systems under consideration,
rearrangement of the Landau state occurs by means of a cascade
of topological transitions, at which the number of sheets of the
Fermi surface grows steadily, as the spectrum of single-particle
excitations $\epsilon(p)$ acquires additional zeroes, in effect becoming
flatter and flatter.  Tracing the evolution of this spectrum as
the coupling constant increases, we have shown that at zero
temperature, the salient feature of the final stage of the evolution is
the formation of the flat bands whose spectrum $\epsilon(p)$
proves to be dispersionless at $T=0$, while at low $T \neq 0$
its dispersion becomes proportional to $T$.

As we have seen in Sec.~III, the underlying reason for such a
universal rearrangement is based on the Pitaevskii identity,
\cite{pit,yaf2001} derived from the Galilean invariance of the Hamiltonian
of the system and gauge invariance, an identity which
coincides with the Landau equation for the spectrum $\epsilon(p)$ as
deduced from the assumption that $\epsilon(p)$ is a functional
of the quasiparticle momentum distribution $n(p)$.  This coincidence
establishes that the ground-state energy $E$ is indeed a functional of
$n$, because, according to the Lehmann expansion, $\epsilon(p,n)$
is a variational derivative of $E$ with respect to $n$. If one
does not care about obedience to the Pauli principle, the global
minimum of the functional $E(n)$ is then attained at some
continuous function $n_*(p)$, devoid (as a rule) of jumps in
momentum space.  But we know that in weakly or moderately
correlated Fermi systems, such solutions violate the Pauli
restriction $n(p)\leq 1$, and, as per standard FL theory,
the true quasiparticle momentum distribution turns out to be
the step function   $\theta(p_F-p)$.

In strongly correlated Fermi systems, as we have seen in Sec.~V.A,
the situation changes drastically: beginning with a critical
coupling constant $\lambda_{\fc}$, the inequality $0 \leq n(p) \leq 1$
is met, lifting the theoretical ban on the emergence of such
an exotic creation as the fermion condensate.

Establishing this freedom does not in itself
 qualify as a rigorous proof of the viability of the swelling
scenario, as such a proof must surmount obstacles associated
with the interplay between single-particle and collective
degrees of freedom.  The flattening of the single-particle
spectrum may lead to softening of some spin/density-fluctuation
mode, in principle generating a corresponding second-order
phase transition before fermion condensation sets in.  Alas,
the detailed circumstances of this interplay await proper
clarification.  Thus it seems that we are forced to seek
guidance from analysis of the available experimental data,
even though, according to Migdal's criterion, these data are
generally of secondary importance.

The flattening phenomena observed experimentally are so pervasive,
however, that an exception to A.B.'s policy statement is in
order.  The discovery of the quantum critical point, made almost
simultaneously in 2D liquid $^3$He, MOSFETs, and heavy-fermion
metals,\cite{godfrin1995,shashrev,lohr,steglich} has been a milestone in
experimental exploration of strongly correlated Fermi systems,
providing a vigorous impetus to their theoretical investigation.
Pivotal experimental guidance to theoretical development has
emerged from evidence for the separation of QCPs from points
of putative second-order phase transitions, uncovered first in
2D liquid $^3$He \cite{saunders2} and more recently in
heavy-fermion metals.\cite{stegcol} Thus separation proves
that collective degrees of freedom are, after all, not of
crucial importance to the QCP phenomena under study.

At this juncture in the development of QCP physics, one is
prompted to ask whether we are facing a situation similar
to that surrounding the discovery of the $W$ boson if in fact
the phenomenon of flattening of single-particle spectra has
already been observed in experiments on condensed-matter systems.
If so, then the misinterpretation of these experiments has impeded
the revelation of a fundamentally new class of Fermi liquids.
Unfortunately, the scope of the data available for sufficiently
definitive analysis is still quite limited.  Measurements of
angle-resolved photoemission electron spectra (ARPES) in solids,
while otherwise promising, are not yet accurate enough
to confirm or refute the linearity in $T$ of the low-temperature
dispersion of these spectra in relevant cases.

Valuable insights might also be drawn from measurements of magnetic
oscillations.  In these measurements, the electron effective mass
is extracted by performing a Lifshitz-Kosevich fit to the temperature
dependence of the magnitude of observed oscillations. In strongly
correlated electron systems of high-$T_c$ superconductors and
heavy-fermion metals possessing a QCP, such analysis is as yet limited
to isolated examples. To our knowledge, a QCP has so far been documented
only in the single high-$T_c$ superconductor YbBa$_2$Cu$_3$O$_{6+x}$,
at a critical doping $x_c\simeq 0.5$ \cite{lonzarich} where
the measurements have been carried out at a temperature around 1 K
in strong magnetic fields $H>55$ T. Additionally, measurements \cite{julian}
performed at much lower temperatures around 20 mK on the heavy-fermion metal
CeCoIn$_5$, which also has a QCP, have revealed deviations from
the Lifshitz-Kosevich formula itself.
Unfortunately, interpretation of the data on magnetic oscillations
is burdened due to the necessity of imposing strong magnetic
fields, which causes substantial distortion of the original
electron motion, triggering  magnetic breakdown.\cite{abrikosov}

In closing, let us briefly revisit the theoretical challenge
presented by the growing body of thermodynamic measurements
on strongly correlated Fermi systems. We focus on the following
striking features of the inferred thermodynamic behavior at
temperatures exceeding what may be interpreted as the critical
temperature $T_m$ for melting of the FL structure and formation
of a flat segment of the quasiparticle spectrum $\epsilon(p)$:
\begin{itemize}
\item[(i)]
The existence of a $T$-independent entropy excess $S_*$, reflected
for example in a huge enhancement upon the FL value of the thermal
expansion coefficient $\beta\simeq \beta_*\propto\partial S_*
/\partial P$.
\item[(ii)]
A sharp falloff of the specific heat $C(T)$ upon passing through the
critical temperature $T_m$.
\item[(iii)]
Curie behavior of the magnetic susceptibility $\chi(T)=C_{\rm eff}/T$, with a effective Curie constant differing from the conventional value.
\end{itemize}

We now comment specifically on each of these behaviors in turn, within
the context of the extension of the Landau-Migdal quasiparticle
theory to embrace topological phase transitions, and especially
fermion condensation.
\begin{itemize}
\item[(i)]
In ordinary FL theory, the entropy $S$ is given by Eq.~(\ref{entr});
the curve $S(T)$ starts at the origin and rises linearly with $T$.
To attain entropy values $S\simeq \ln 2$ at low $T$, the density of
states, which specifies the slope $dS(T)/dT$, must be further
enhanced beyond the standard heavy-fermion boost associated with
the width of the narrow $f$-band lying exactly at the Fermi
surface. Nevertheless, experimental measurements \cite{hossain}
indicate that the entropy value $S=0.5 \ln 2$ is realized in the
heavy-fermion metal YbIr$_2$SI$_2$ at $T>T_m\simeq 1~K$.  Moreover,
another apparent manifestation of the flattening phenomena is exhibited
by the heavy-fermion metal CeCoIn$_5$ mentioned above.  Its thermal
expansion coefficient $\beta(T)$, measured in external magnetic
fields $H\simeq 5T$ imposed to suppress superconductivity, increases
strongly until $T$ reaches the value $T_m=0.3\,K$,\cite{steg2007}
whereas $T>T_m$, $\beta(T)$ becomes $T$-independent, being
enhanced by a factor $10^3-10^4$ compared with standard FL values.

\item[(ii)]
A drop of the Sommerfeld ratio $\gamma(T)$ is observed in many
strongly correlated electron systems as a temperature $T_m$ is
exceeded.  The record slump is found in YbIr$_2$Si$_2$ where the FL contribution to $\gamma(T)$ {\it collapses} at $T>T_m=0.7\,K$. Usually such a slump is attributed to some second-order (antiferromagnetic) phase transition. (As is well known, in
the Landau theory of second--order phase transitions there is a jump of the specific heat at the transition point.)  Significantly, the experimental data on this compound and any other compounds having the QCP fail to show any jump of $C(T)$ at any $T$. The scaling theory of second-order phase transitions also fails to explain the experimental behavior of $C(T)$. In addition, numerous attempts to establish the structure of the corresponding order parameter on the side of $T_m$
allegedly associated with the ordered phase, have also been unsuccessful. In light of all that has been done before in this article, none of this comes as a big surprise, and reminds us of a saying by Confucius:

{\it It is hard to find a black cat in a dark room, especially if it
is not there}.

Indeed, the arguments given in Sec.~V.E explain the behavior of
$C(T)$ in the vicinity of $T_m$, attributing it to a {\it crossover}
from a state with a multi-connected Fermi surface to a state having a
{\it flat band}.  Bearing in mind this association, we infer
that in the QCP region, there are no hidden order parameters at all,
since the behavior stems from topological phase transitions
rather than second-order, symmetry-breaking phase transitions.

\item[(iii)]
Curie-like behavior of the magnetic susceptibility $\chi(T>T_m)=C_{\rm eff}/T$,
which is incompatible with the customary FL Pauli behavior $\chi(T)={\rm const}$,
was first observed in 2D liquid $^3$He.\cite{godfrin1998}  An analogous
Curie-like behavior of $\chi(T)$ has also been seen recently at very
low temperatures in normal states of high-$T_c$ superconductors
placed into a strong magnetic field to suppress the superconductivity.\cite{loram1,loram2} In both the cases, the effective Curie constant
$C_{\rm eff}$, being nontrivially dependent on the density $\rho$ or doping $x$,
is at variance with the ordinary Curie constant proportional to $\rho$.
\end{itemize}

A detailed discussion of experimental support for or against
the topological scenario for the quantum critical point is
beyond the scope of this article, especially bearing in mind
that the measurements involved are for the most part very fresh,
and---as cautioned by A.\ B.\ Migdal---may be subject to error
or incomplete analysis.  Thus, in summarizing the current
 state of knowledge,
we are forced to recognize that the envisioned
swelling of the Fermi surface in
most strongly correlated condensed-matter many-fermion
systems featuring the occurrence of the flat bands,
has been neither validated nor disproved experimentally.
The fermion condensate remains as elusive as the Cheshire Cat
of {\it Alice in Wonderland}, teasing us with mischievous grins
that become more and more visible and numerous.

We gratefully acknowledge discussions with A.\ Alexandrov,
H.\ Godfrin, V.\ Shaginyan, F.\ Steglich and G.\ Volovik.  This
research was supported by the McDonnell Center for the Space Sciences,
by Grants No.~2.1.1/4540 and NS-7235.2010.2 from the Russian
Ministry of Education and Science, and by Grant No.~09-02-01284
from the Russian Foundation for Basic Research.

\end{document}